%% file: paper.tex
\PassOptionsToPackage{table}{xcolor}
\documentclass[sigconf]{acmart}

\usepackage{hyperref}
\usepackage{booktabs}
\usepackage{multirow}
\usepackage{graphicx}
\usepackage{enumitem}
\usepackage{subfigure}  
\usepackage{graphicx}
\usepackage{bbding} 
\usepackage{stfloats}
\usepackage{float}
\usepackage{makecell}
\usepackage{algorithm} 
\usepackage{microtype}
\usepackage{algpseudocode}
\usepackage[title]{appendix}
\usepackage{bbding}
\usepackage{balance} 
\usepackage{tabularx}
\usepackage{array}
\usepackage{ragged2e}

\definecolor{headergray}{RGB}{245,245,245}
\definecolor{rowgray}{RGB}{240,240,240}
\definecolor{impgreen}{RGB}{0,100,0} 
\definecolor{mutedred}{RGB}{176,60,60}
\definecolor{lightgray}{gray}{0.95}

\definecolor{oursgray}{RGB}{238,238,238}
\definecolor{deltagray}{RGB}{248,248,248}
\definecolor{gainred}{RGB}{150,70,55}

\title{\textsc{OneLatent}: Latent Reasoning for Efficient Foundation Recommendation Models}

\author{Hao Jiang\textsuperscript{*},
Peiru Du\textsuperscript{*},
Pengfei Yao\textsuperscript{*},
Mengting Li,
Siyuan Lou,
Kuo Cai,
Sheng Yu,
Qiang Luo,
Jian Liang,
Ruiming Tang,
Fei Pan\textsuperscript{\textdaggerdbl},
Peng Jiang\textsuperscript{\textdaggerdbl},
Wenwu Ou\textsuperscript{\textdaggerdbl}}
\affiliation{%
  \institution{Kuaishou Technology}
  \city{Beijing}
  \country{China}
}
\email{{jianghao11,dupeiru,yaopengfei03,limengting03,lousiyuan05,
caikuo,yusheng03}@kuaishou.com}
\email{{luoqiang,liangjian03,tangruiming,panfei05,jiangpeng,
luocheng10}@kuaishou.com}


\begin{document}

\begin{abstract}
Large language models (LLMs) have demonstrated strong reasoning capabilities, motivating their adoption as the backbone of foundation recommendation models (FRMs). Existing approaches typically enhance recommendation performance by eliciting explicit Chain-of-Thought (CoT) reasoning under the \textbf{Think-then-Answer} paradigm. However, explicit CoT in FRMs incurs substantial inference overhead by generating lengthy reasoning traces. Moreover, it relies on manually designed templates that struggle to capture diverse and dynamic user interests in real-world scenarios. To address these limitations, we propose \textsc{OneLatent}, an efficient latent reasoning framework for FRMs. Its key idea is to compress explicit reasoning traces into several learnable latent reasoning tokens, thereby enabling efficient \textbf{Latent-Reason-then-Answer} inference without generating verbose reasoning traces. Specifically, \textsc{OneLatent} first introduces Multi-View Adaptive CoT (MV-ACoT), which constructs diverse, high-quality, teacher-generated CoT supervision by exploring user interests from multiple perspectives while automatically adapting reasoning complexity to each instance. Building on pretrained FRMs, \textsc{OneLatent} then employs a three-stage latent token alignment paradigm to progressively internalize teacher-generated CoT traces into learnable latent tokens. Finally, a multi-stage, curriculum-based post-training strategy effectively activates latent token reasoning for downstream recommendation tasks. Extensive experiments on an industrial-scale Kuaishou dataset and the public Kuaishou LLM-Rec benchmark demonstrate that \textsc{OneLatent} consistently outperforms both explicit CoT-based methods and traditional baselines. Compared with the \textit{Think} and \textit{No-Think} variants of FRMs, \textsc{OneLatent} improves SID@64 by 17.44\% and 9.33\%, respectively, while achieving more than $17.49\times$ higher online inference throughput. To facilitate real-world deployment, we develop a production serving system that enables full-traffic deployment of \textsc{OneLatent} in production environments with scalable, real-time inference. An online A/B test in Kuaishou's local-services advertising scenario shows an estimated 9.6\% revenue lift over strong online baselines, including OneRec and OneReason.
\end{abstract}

\maketitle

\begingroup
\renewcommand{\thefootnote}{}
\footnotetext{\textsuperscript{*} Equal contribution. \textsuperscript{\textdaggerdbl} Corresponding authors.}
\endgroup

\begin{figure}[htp]
  \centering
  \includegraphics[ width=\linewidth, height=4cm]{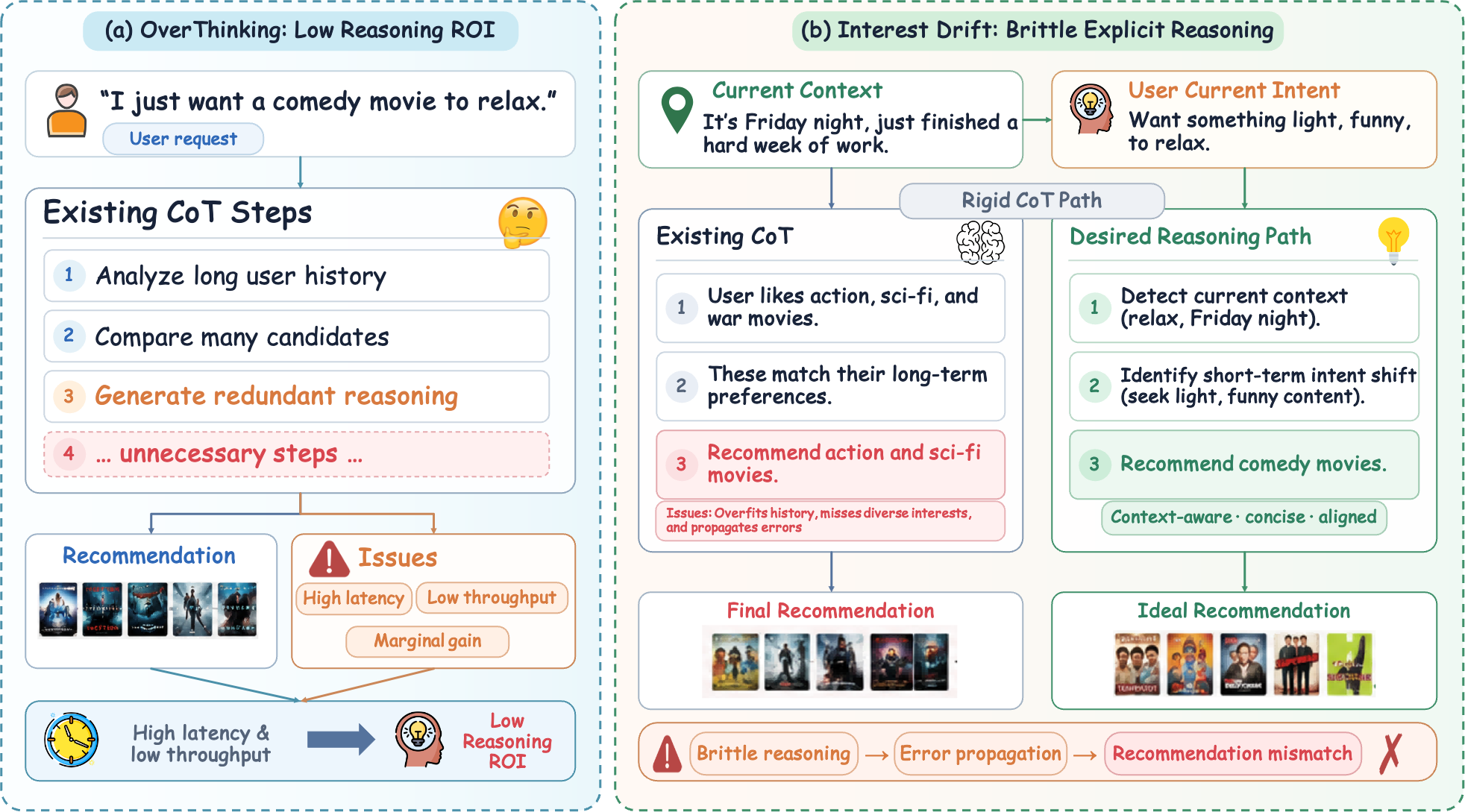}
  \caption{Limitations of explicit CoT for recommendation: overthinking with low reasoning ROI and interest drift.}
  \Description{Two examples illustrate redundant explicit reasoning for a simple request and interest drift caused by a rigid reasoning path.}
  \label{intro}
\end{figure}

\section{Introduction}
Driven by the scaling of model capacity and data, LLMs have progressed from generation to knowledge-intensive reasoning and agentic decision making~\cite{jaech2024openai, guo2025deepseek}. This progress has motivated researchers to transfer these capabilities to recommender systems by adapting language-model architectures, or even language models themselves, to improve existing recommendation systems~\cite{zhai2024actions, zhang2024wukong, rajput2018recommender}. Recent generative recommendation models, exemplified by the \textsc{OneRec} family~\cite{zhou2025onerec, zhou2025onerecv2, liu2025onerecthink, zhou2025openonerec}, have demonstrated that Transformer-based architectures can scale recommendation modeling to support industrial-scale end-to-end deployment. However, they mostly make predictions through implicit representation computing, making their decision processes difficult to interpret or control~\cite{zhou2017deep, zhou2019deep}. Although several studies introduce latent reasoning mechanisms into recommendation~\cite{tang2026think, dai2025onepiece}, their reasoning still operates in a non-semantic latent space, making it difficult to determine whether the learned representations encode coherent, task-relevant reasoning.

A more intuitive approach is to directly use LLMs as the backbone of foundation recommendation models (FRMs) and adapt them through recommendation-specific training~\cite{yi2025recgpt, yi2025recgptv2, zhou2025openonerec}. Recent work, particularly OneReason~\cite{team2026onereason}, demonstrates the success of the {Think-then-Answer} paradigm for FRMs, in which the model reasons over user preferences and context before generating recommendations. However, recommendation reasoning cannot simply inherit generic CoT patterns~\cite{zhou2025perception, zhang2026thinking}, as user interests are diverse and evolving, requiring structured reasoning from preference induction to intent inference and decision making. Despite their success, explicit CoT-based FRMs face two key limitations.

\textbf{(1) Overthinking and low reasoning ROI.}
As shown in Figure~\ref{intro}(a), explicit CoT often generates lengthy and redundant reasoning even for simple requests with clear intents. This increases inference latency and reduces throughput, while bringing only marginal recommendation gains.

\textbf{(2) Interest drift and brittle explicit reasoning.}
As shown in Figure~\ref{intro}(b), a fixed, single-path explicit-CoT strategy can overemphasize dominant historical preferences while overlooking short-term, context-dependent interests. If even a minor error or hallucination occurs during reasoning, it may propagate to the final output and result in a mismatched recommendation.

To address these limitations, we propose \textsc{OneLatent}, a latent-reasoning framework for FRMs. Rather than autoregressively emitting a rationale for every recommendation decision, \textsc{OneLatent} inserts three special learnable latent tokens between the user context and the target SID, enabling a \textbf{Latent-Reason-then-Answer} interface with constant reasoning-token overhead. Specifically, \textsc{OneLatent} introduces MV-ACoT, which constructs complementary teacher supervision from multiple perspectives and prompts the teacher model to adapt its reasoning depth to the complexity of the observed behavioral signals. The latent token alignment procedure consists of three stages. It first learns contextual representations at the latent-token positions through rationale prediction and then connects the same latent interface to next-item prediction. Finally, curriculum-based post-training supports both direct and latent-reasoning modes for downstream recommendation. Because the acceleration provided by \textsc{OneLatent} alone cannot meet the latency requirements of synchronous inference at full traffic scale, we further deploy a production serving architecture that prioritizes valuable asynchronous updates under limited GPU capacity while keeping foundation-model inference off the online request path.

We conduct experiments on an industrial Kuaishou recommendation dataset and the public Kuaishou LLM-Rec benchmark. Our primary \textsc{OneLatent} model is built on Qwen3.5-0.8B\footnote{\url{https://huggingface.co/Qwen/Qwen3.5-0.8B}} and outperforms all compared methods across the reported metrics. On the industrial dataset, it improves SID@64 by 17.44\% over the explicit \textit{Think} variants and by 9.33\% over its \textit{No-Think} variants, while delivering a $17.49\times$ throughput gain over explicit rationale decoding. We further train and evaluate \textsc{OneLatent} based on Qwen3.5-9B\footnote{\url{https://huggingface.co/Qwen/Qwen3.5-9B}} and find that it also outperforms both direct prediction and explicit CoT baselines built on the same 9B backbone. Analyses of MV-ACoT quality, adaptive reasoning length, and view-conditioned semantic reconstruction provide complementary evidence for the proposed training recipe, while we explicitly distinguish semantic association from a strict information bottleneck claim. The online A/B test against strong online baselines, including OneRec and OneReason, further validates the effectiveness of real-time \textsc{OneLatent} and indicates an estimated 9.6\% revenue lift in Kuaishou’s local-services advertising scenario.

Our main contributions are summarized as follows:
\begin{itemize}[leftmargin=10pt,topsep=0pt]

\item We propose \textsc{OneLatent}, an efficient latent reasoning framework that introduces the \textbf{Latent-Reason-then-Answer} paradigm for FRMs and improves performance.

\item We develop \textbf{MV-ACoT} for high-quality, multi-view CoT supervision, together with a three-stage latent-token alignment procedure and curriculum-based post-training that learn a latent-token-conditioned interface from explicit rationale supervision.

\item We build a \textbf{Production Serving System} for FRMs that controls inference cost through behavior-triggered admission, capacity-aware scheduling, and asynchronous result materialization, enabling \textbf{Full-Traffic Deployment} of \textsc{OneLatent} in production environments. An online A/B experiment indicates an estimated 9.6\% revenue lift over a strong online baseline in Kuaishou's local-services advertising scenario.
\end{itemize}

\section{Related Work}

\subsection{Scaling Recommendation with LLMs}
Traditional recommendation methods learn user representations from historical interactions and predict the next item~\citep{hidasi2015session, kang2018self, sun2019bert4rec}. With the rise of LLMs, early LLM-based approaches directly prompted LLMs to make recommendation decisions~\citep{geng2022recommendation, gao2023chat, hou2024large}. Although these methods leverage the world knowledge of LLMs, they remain limited in modeling collaborative signals.
Recent studies have advanced recommendation systems toward foundation recommendation models along two complementary directions. The first develops recommendation-native backbones: TIGER represents items using semantic IDs~\citep{rajput2023recommender}, HSTU formulates large-scale recommendation as sequence transduction~\citep{zhai2024actions}, and OneRec unifies retrieval and ranking in an end-to-end generative architecture~\citep{deng2025onerec}. The second adapts pretrained LLMs to recommendation data. LC-Rec aligns linguistic knowledge with collaborative semantics~\citep{zheng2024adapting}, while the RecGPT family develops transferable LLM-based recommenders through recommendation-domain pre-training~\citep{yi2025recgpt, yi2025recgptv2}. Despite their success, these models mainly focus on behavior modeling and item generation, providing limited supervision for inferring latent user needs and making recommendation decisions.

\subsection{Reasoning-Enhanced Recommendation}
Recent work has begun to equip recommendation models with reasoning capabilities. Latent reasoning methods reduce inference costs by avoiding lengthy natural-language rationales. ReaRec recursively refines user representations in the hidden space~\citep{tang2026think}, OnePiece introduces reasoning modules into industrial ranking models~\citep{dai2025onepiece}, OneSearch-V2 distills LLM reasoning into compact internal states~\citep{chen2026onesearch}, and REG4Rec explores multiple latent reasoning paths through self-reflection~\citep{xing2025reg4rec}.
In parallel, explicit reasoning methods expose intermediate reasoning processes in natural language. GOT4Rec~\citep{long2024got4rec}, Reason4Rec~\citep{fang2025reason4rec}, and ThinkRec~\citep{yu2026thinkrec} construct user preference analyses or multi-path reasoning trajectories through supervised fine-tuning. RecOne~\citep{kong2026think}, OneRec-Think~\citep{liu2025onerec}, and OpenOneRec~\citep{zhou2025openonerec} further combine CoT supervision with reinforcement learning to improve recommendation decisions. OneReason~\citep{team2026onereason} reports that its thinking mode does not consistently outperform non-thinking inference. PauseRec~\citep{he2026implicit} further identifies weakened verbalization, text--SID representation mismatch, and rationale sensitivity as limitations of explicit reasoning training, and replaces rationales with implicit pause tokens.
Although these methods advance reasoning-based recommendation, explicit reasoning is costly and sensitive to the interface between natural language and item IDs, whereas latent reasoning often lacks structured recommendation supervision. Our approach uses complementary multi-view rationales to train a latent-token-conditioned interface and employs staged alignment for SID prediction.

\section{Methodology}

\begin{figure*}[tp]
  \centering
    \vspace{-3mm}
  \includegraphics[width=\textwidth,     height=6.5cm
  ]{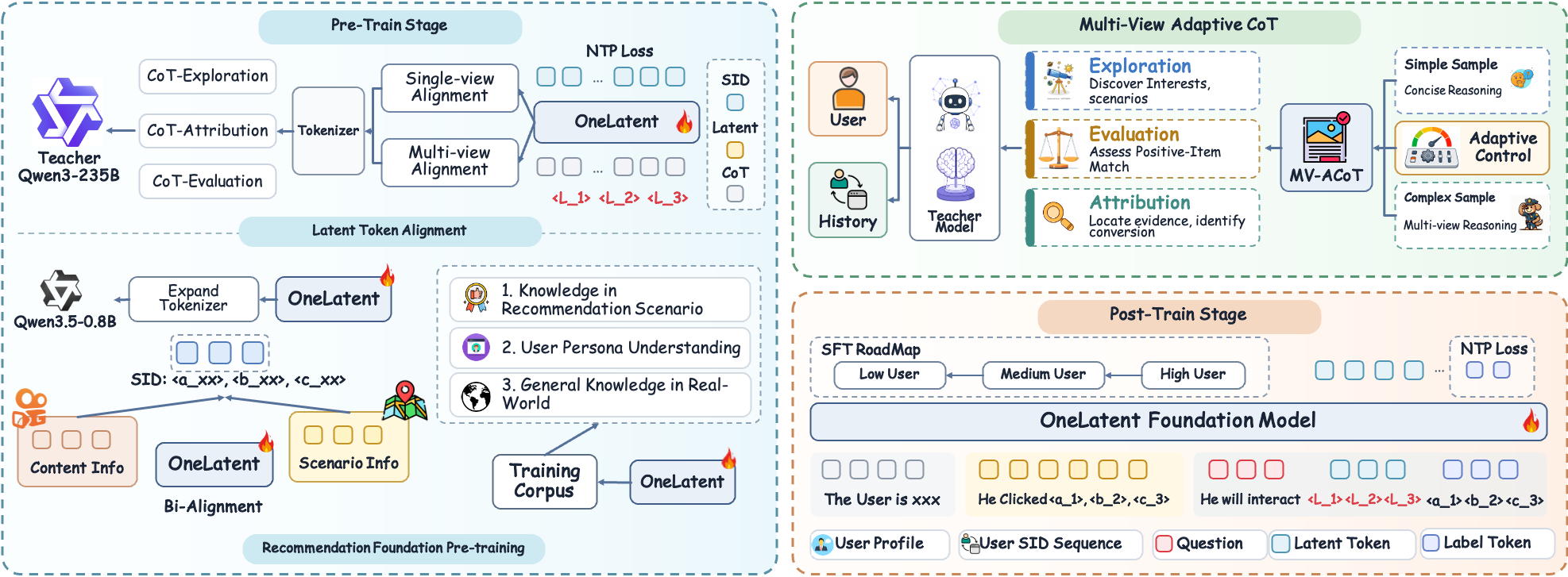}
  \caption{Overview of \textsc{OneLatent}, comprising Multi-View Adaptive CoT, three stage latent-token alignment, and curriculum-based post-training.}
  \Description{The framework combines recommendation foundation pretraining, multi-view adaptive CoT supervision, two stages of rationale alignment, recommendation-oriented context alignment, and curriculum-based post-training.}
  \label{fig:framework}
\end{figure*}

\subsection{Problem Formulation and Framework}

For a user $u$, let $P_u$ denote the user profile, $H_u$ the chronologically ordered interaction history, $\mathcal{C}_u$ the request-time context observed before prediction, and $i_u^+$ the next interacted item. The context $\mathcal{C}_u$ may include spatiotemporal information and user- or request-specific textual signals, such as search queries, comments, and private messages. Each item $i$ is represented by a three-token Semantic ID (SID), $\mathcal{S}i=[s_i^{(1)};s_i^{(2)};s_i^{(3)}]$, constructed using Res-Kmeans~\cite{luo2024qarm}. The recommendation task is to predict $\mathcal{S}{i_u^+}$ conditioned on $(P_u,H_u,\mathcal{C}_u)$. To prevent target leakage, $i_u^+$ and its SID are used only as supervision and are never provided as model inputs.

We add three special latent token IDs to the FRM vocabulary and denote their shared sequence by
\begin{equation}
\mathbf{L}=
[\langle L_1\rangle;\langle L_2\rangle;\langle L_3\rangle].
\label{eq:latent-token-ids}
\end{equation}
These are fixed token IDs shared by all samples, rather than sample-specific latent vectors. When the Transformer processes a sequence containing $\mathbf{L}$, it produces contextualized hidden representations at these token positions. Subsequent rationale or SID tokens attend to these representations through standard causal self-attention. We use $p_{\mathrm{T}}$ for the teacher LLM and $p_\phi$ for the fully trainable student FRM. We reserve $\mathcal{L}$ for training losses.

As shown in Figure~\ref{fig:framework}, \textsc{OneLatent} comprises three components. First, Multi-View Adaptive Chain-of-Thought (MV-ACoT) constructs complementary teacher-generated supervision for intent exploration, positive-item evaluation, and interaction attribution. Second, Latent Token Alignment trains the contextualized hidden representations induced at the latent-token positions to support rationale generation. Third, curriculum-based post-training jointly optimizes standard recommendation and latent-reasoning modes. During online inference, the model takes the user context followed by the three latent token IDs and directly generates the target SID, avoiding autoregressive natural-language rationale generation.

\subsection{Multi-View Adaptive CoT}
\label{sec:Multi-View Adaptive CoT}
In this section, we propose MV-ACoT to construct high-quality, recommendation-oriented CoT supervision that equips FRMs with latent reasoning capabilities.

\subsubsection{Multi-View CoT Task Design}
Effective recommendation CoT should expose decision-relevant signals without imposing the same reasoning template on every request. MV-ACoT therefore constructs complementary \textbf{Exploration}, \textbf{Evaluation}, and \textbf{Attribution} supervision. The views differ in the input information available to the teacher model, but their outputs are used only as student targets during latent alignment.

\paragraph{View 1: Exploration}
The exploration teacher receives only the user profile and interaction history, without a candidate or target item. It generates both a textual rationale and a list of plausible commercial intents:
\begin{equation}
(R_u^{\mathrm{exp}},\mathcal{I}_u)
\sim
p_{\mathrm{T}}\!\left(
\cdot \mid \pi_{\mathrm{exp}}(P_u,H_u)
\right),
\label{eq:exploration}
\end{equation}
where $R_u^{\mathrm{exp}}$ is the rationale and $\mathcal{I}_u$ is the intent list. This view supports open-ended demand discovery and interest-shift analysis.

\paragraph{View 2: Evaluation}
The evaluation teacher receives the observed positive item $i_u^+$ as privileged guidance and assesses its match to the user evidence. We use only the real positive target and do not introduce sampled negative candidates. The teacher produces a textual rationale and a discrete matching score $q_u^{\mathrm{eval}}\in\{1,\ldots,5\}$:
\begin{equation}
(R_u^{\mathrm{eval}},q_u^{\mathrm{eval}})
\sim
p_{\mathrm{T}}\!\left(
\cdot \mid
\pi_{\mathrm{eval}}(P_u,H_u,\mathcal{C}_u,i_u^+)
\right),
\label{eq:evaluation}
\end{equation}
where a larger score indicates stronger user--item consistency. Both the rationale and score are retained as supervision.

\paragraph{View 3: Attribution}
The attribution teacher is explicitly told that $i_u^+$ is a positive interaction and explains which preceding signals can account for it:
\begin{equation}
R_u^{\mathrm{attr}}
\sim
p_{\mathrm{T}}\!\left(
\cdot \mid
\pi_{\mathrm{attr}}(P_u,H_u,\mathcal{C}_u,i_u^+)
\right),
\label{eq:attribution}
\end{equation}
This view outputs only the textual attribution rationale. It does not add a category, confidence value, or matching score. For notational convenience, let $Y_u^{(v)}$ denote the complete teacher-generated output for view $v$, where $Y_u^{(\mathrm{exp})}=[R_u^{(\mathrm{exp})};\mathcal{I}_u]$, $Y_u^{(\mathrm{eval})}=[R_u^{(\mathrm{eval})};q_u^{(\mathrm{eval})}]$, and $Y_u^{(\mathrm{attr})}=R_u^{(\mathrm{attr})}$. \textsc{Exploration} does not use target information, whereas \textsc{Evaluation} and \textsc{Attribution} expose the positive item only to the teacher. This information serves as privileged supervision rather than student input: the item description, target SID, and candidate set are excluded from the student prefixes in Stages I/II.

\subsubsection{Adaptive CoT Generation}

Beyond multi-view CoT construction, MV-ACoT uses a prompt-guided, self-adaptive reasoning policy. It does \emph{not} employ a separately trained difficulty classifier, a Low/High label, or a hand-set decision threshold. Instead, each view prompt instructs the teacher to first assess the complexity of the evidence in $P_u$ and $H_u$, with particular attention to strong behaviors, conflicting interests, weakly related events, and noise, and then calibrate the reasoning path accordingly:
\begin{equation}
Y_u^{(v)} \sim p_{\mathrm{T}}\!\left(
\cdot \mid \pi_v^{\mathrm{adapt}}(G_u^{(v)})
\right),
\end{equation}
where $G_u^{(v)}$ denotes the information available to teacher view $v$ as specified above. For simple cases with a dominant signal, the prompt asks the teacher to converge quickly and summarize secondary noise. For complex cases, it asks for multi-dimensional comparison. Reasoning dimensions such as geography, time, signal strength, industry constraints, and conversion stage are selected only when supported by the instance. The teacher does not emit a difficulty label; adaptation is reflected directly in the content and length of $Y_u^{(v)}$. Appendix~\ref{app:mv_acot_prompt} provides the prompt principles.

\subsection{Latent Token Alignment}
Starting from the pretrained FRM $\mathcal{F}_{\mathrm{base}}$, we perform training in three sequential stages: single-view rationale alignment, multi-view rationale alignment, and recommendation-oriented context alignment. In Stages I and II, the model learns to reproduce teacher supervision following the three latent token IDs. In Stage III, the textual supervision target is replaced with the SID of the next item. Notably, our \textsc{OneLatent} framework introduces no additional encoder, projector, or decoder modules.

\subsubsection{FRMs Pre-training}
We initialize from Qwen3.5-0.8B and extend its vocabulary with Semantic ID tokens. Each item $v$ is represented by a three-level SID sequence with Res-Kmeans~\cite{luo2024qarm}:
\begin{equation}
\mathcal{S}_v =
[s_v^{(1)}; s_v^{(2)}; s_v^{(3)}],
\end{equation}
where $s_v^{(l)}$ denotes the SID at level $l$, encoding item-side multimodal information~\cite{luo2025qarm}. We bidirectionally align $\mathcal{S}_v$ with the item's dense caption $C_v$:
\begin{equation}
\mathcal{L}_{\mathrm{sem}}
=
-\log p_{\phi}(\mathcal{S}_v \mid C_v)
-\log p_{\phi}(C_v \mid \mathcal{S}_v).
\end{equation}
Moreover, we introduce diverse CPT tasks to enhance the FRMs, including recommendation-domain knowledge such as spatiotemporal context, user persona understanding, and general-domain knowledge. Together, these tasks construct a capable foundation model for recommendation.

\subsubsection{Stage I: Single-View Latent Reasoning Alignment}

We first warm up the latent-token-conditioned interface using one retained teacher output for each training context. Let $X_u=[P_u;H_u]$, let $v_u^\star$ denote the selected reasoning view, and let $\langle v_u^\star\rangle$ be its control token. The student sequence is
\begin{equation}
[X_u;\langle v_u^\star\rangle;\mathbf{L};
\widetilde{Y}_u^{(v_u^\star)}].
\end{equation}
where $\widetilde{Y}$ denotes the retained teacher output, including the auxiliary intent list or score when required by the view. We apply loss only to these output tokens:
\begin{equation}
\mathcal{L}_{\mathrm{single}}
=
-\mathbb{E}_{u}
\sum_{t=1}^{|\widetilde{Y}_u^{(v_u^\star)}|}
\log p_\phi\!\left(
\widetilde{y}_{u,t}^{(v_u^\star)}
\mid
X_u,\langle v_u^\star\rangle,\mathbf{L},
\widetilde{y}_{u,<t}^{(v_u^\star)}
\right).
\label{eq:single}
\end{equation}
The view token specifies the type of rationale to generate. Conditioned on this view, the following three latent tokens compress the user context into compact representations that the FRM can use for rationale generation.

\subsubsection{Stage II: Multi-View Latent Reasoning Alignment}

We then use every retained MV-ACoT output. Let
$\mathcal{V}_u\subseteq
\{\mathrm{exp},\mathrm{eval},\mathrm{attr}\}$ be the available views for user
$u$. Let
$C_{u,t}^{(v)}
=(X_u,\langle v\rangle,\mathbf{L},\widetilde{y}_{u,<t}^{(v)})$
denote the decoding context at step $t$. The multi-view objective is
\begin{equation}
\begin{aligned}
\ell_u^{(v)}
&=
-\frac{1}{|\widetilde{Y}_u^{(v)}|}
\sum_{t=1}^{|\widetilde{Y}_u^{(v)}|}
\log p_\phi\!\left(
\widetilde{y}_{u,t}^{(v)}
\mid C_{u,t}^{(v)}
\right),\\
\mathcal{L}_{\mathrm{multi}}
&=
\mathbb{E}_{u}\!\left[
\frac{1}{|\mathcal{V}_u|}
\sum_{v\in\mathcal{V}_u}
\ell_u^{(v)}
\right].
\end{aligned}
\label{eq:multi}
\end{equation}
Each output is paired with a view token indicating \textsc{Exploration}, \textsc{Evaluation}, or \textsc{Attribution}. We average token losses within each output so that longer traces do not automatically receive greater weight. For every view, the student prefix is exactly $[P_u;H_u;\langle v\rangle;\mathbf{L}]$. It contains no candidate description, positive-item description, or target SID, even when the teacher used the positive item to construct evaluation or attribution supervision.

\paragraph{Attention and optimization scope.}
All stages use a standard causal attention mask and update all student parameters. Consequently, each output token can attend to both the raw student context $X_u$ and the preceding latent-token positions. We therefore neither impose a strict attention bottleneck nor claim that the alignment objective forces all rationale information to be encoded exclusively in $\mathbf{L}$. Instead, the objective encourages the model to use a shared interface conditioned on the learnable latent tokens across different views. The semantic reconstruction analysis in Section~\ref{sec:latent-interpretability} evaluates the resulting association.

\subsubsection{Stage III: Recommendation-Oriented Context Alignment}

The preceding stages optimize view-specific supervision but do not directly optimize recommendation decisions. We therefore connect the same latent-token-conditioned interface to target-item prediction. In addition to the user profile $P_u$ and historical SID sequence $H_u$, we use request context $\mathcal{C}_u$, including temporal, geographical, and user- or request-bound textual signals. The target SID sequence remains excluded from the input:
\begin{equation}
X_u^{\mathrm{ctx}} =
[P_u; H_u; \mathcal{C}_u;\mathbf{L}].
\end{equation}
The model predicts the observed target SID sequence $\mathcal{S}_u^+$:
\begin{equation}
\mathcal{L}_{\mathrm{align}}
=
-\mathbb{E}
\left[
\log p_{\phi}
\left(
\mathcal{S}_u^+ \mid X_u^{\mathrm{ctx}}
\right)
\right].
\end{equation}
This stage yields the context-aware latent reasoning FRM $\mathcal{F}_{\mathrm{latent\text{-}ctx}}$.

\subsection{Post-Training and Real-Time Inference}

In this section, we further adapt $\mathcal{F}_{\mathrm{latent\text{-}ctx}}$ to the target recommendation scenario through multi-stage curriculum-based recommendation SFT and latent reasoning SFT.

\subsubsection{Multi-stage Curriculum-based Recommendation SFT}
The primary objective of this stage is to equip \textsc{OneLatent} with recommendation capabilities in the target scenario.
We first train the model to predict the next target SID sequence from standard recommendation inputs:
\begin{equation}
X_u^{\mathrm{std}} = [P_u; H_u;\mathcal{C}_u].
\end{equation}
Here, $P_u$ provides persistent user attributes, while $H_u$ captures behavioral preferences. The training objective is
\begin{equation}
\mathcal{L}_{\mathrm{SFT}}
=
-\mathbb{E}
\left[
\log p_{\phi}
\left(
\mathcal{S}_u^+ \mid X_u^{\mathrm{std}}
\right)
\right].
\end{equation}
To smoothly equip an LLM with recommendation capabilities across industrial scenarios with varying levels of user activity, we introduce a {multi-stage curriculum-learning strategy}. Specifically, we employ a curriculum from high-activity to low-activity users:
\begin{equation}
\mathcal{D}_{\mathrm{high}}
\rightarrow
\mathcal{D}_{\mathrm{medium}}
\rightarrow
\mathcal{D}_{\mathrm{low}}.
\end{equation}

\subsubsection{Latent Reasoning-based Recommendation SFT}

We then activate latent reasoning by appending the three latent token IDs to the same user context:
\begin{equation}
X_u^{\mathrm{LR}} =
[P_u; H_u;\mathcal{C}_u;\mathbf{L}].
\end{equation}
The model is optimized to predict the target SID sequence immediately after the latent tokens:
\begin{equation}
\mathcal{L}_{\mathrm{LR}}
=
-\mathbb{E}
\left[
\log p_{\phi}
\left(
\mathcal{S}_u^+ \mid X_u^{\mathrm{LR}}
\right)
\right].
\end{equation}

To preserve performance in both latent-reasoning and standard recommendation modes, we mix latent reasoning and conventional SFT samples at a $1{:}1$ ratio:
\begin{equation}
\mathcal{L}_{\mathrm{post}}
=
\frac{1}{2}\mathcal{L}_{\mathrm{LR}}
+
\frac{1}{2}\mathcal{L}_{\mathrm{SFT}}.
\end{equation}
This objective enables the \textbf{Latent-Reason-then-Answer} paradigm when $\mathbf{L}$ is provided, while retaining stable recommendation performance without latent tokens.

\subsubsection{Latent-Reasoning Inference}
Online inference uses $X_u^{\mathrm{LR}}$ and decodes only the three-level target SID:
\begin{equation}
\widehat{\mathcal{S}}_u
=
\operatorname*{arg\,max}_{\mathcal{S}}
p_\phi(\mathcal{S}\mid P_u,H_u,\mathcal{C}_u,\mathbf{L}).
\label{eq:inference}
\end{equation}
The MV-ACoT teacher, view tokens, privileged item information, and textual rationale are absent at inference. The same autoregressive FRM directly decodes the SID after the three latent token IDs, so reasoning-token overhead is fixed rather than proportional to a generated natural-language trace.

\begin{figure}[t]
  \centering
  \includegraphics[width=\columnwidth]{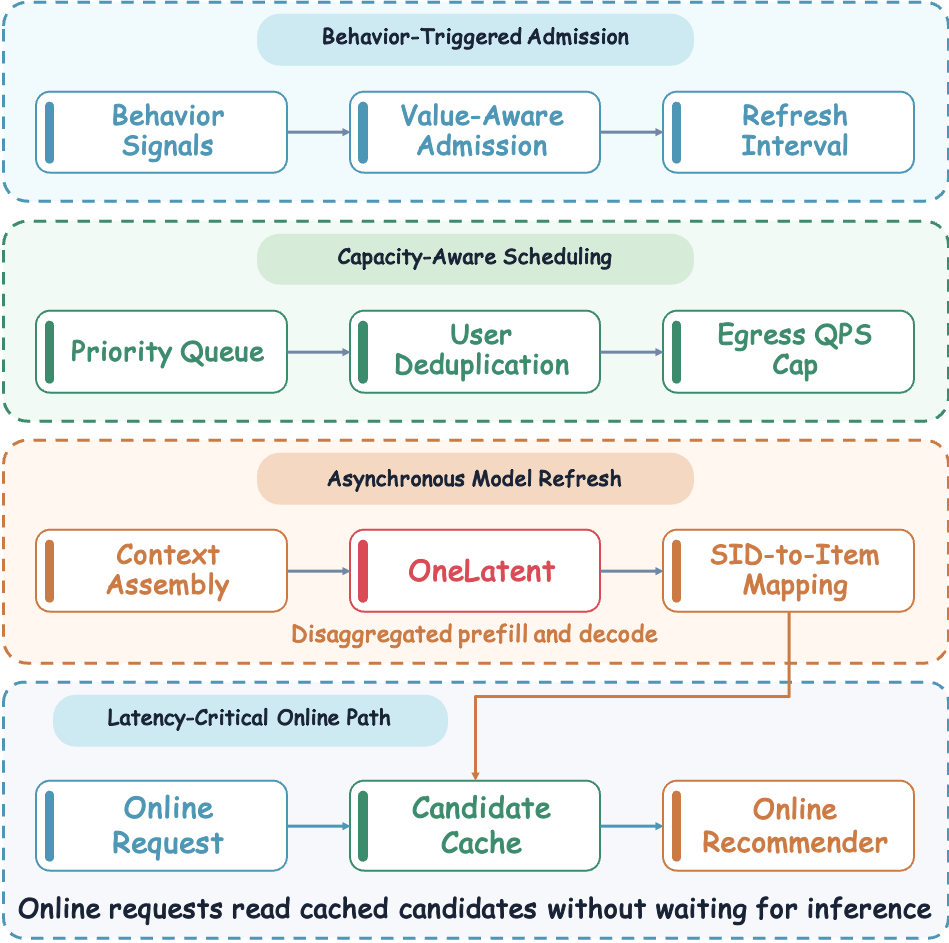}
  \caption{Production online inference architecture.}
  \Description{Behavior-triggered admission and capacity-aware scheduling feed asynchronous OneLatent inference, whose generated candidates are cached for the latency-critical online path.}
  \label{fig:production-serving}
\end{figure}

\subsubsection{Real-Time Production Serving}
\label{sec:production-serving}

Although latent reasoning reduces decoding cost, synchronously invoking the FRM for every request still exceeds production latency and throughput budgets. 

Figure~\ref{fig:production-serving} shows our online inference architecture. 
Following the production serving architecture of OneReason~\cite{team2026onereason}, we deploy the FRM as a dedicated retrieval channel rather than replacing the existing retrieval pipeline. Its candidates are merged with those from the incumbent retrieval pipeline and compete in the same fine-ranking stage, allowing \textsc{OneLatent} to complement the established recommendation system. We extend this architecture with behavior-triggered real-time inference. Streaming interactions undergo value-aware admission, with update priorities and refresh intervals determined by user value and behavior strength. The scheduler deduplicates requests, preserves the latest context, prioritizes high-value updates, and enforces a strict egress QPS cap. For each admitted update, an asynchronous gateway constructs $P_u;H_u;\mathcal{C}_u;\mathbf{L}$, invokes the disaggregated gRPC inference service, maps generated SIDs to ranked items, and caches the results in Redis. The online system retrieves these cached candidates and retains the previous result when an update is delayed or unavailable. Consequently, recent behavior can promptly influence ranking while model inference remains outside the latency-critical request path and GPU load remains bounded.

\section{Experiments}

We organize the experiments around five questions:
\begin{itemize} [leftmargin=7pt,topsep=0pt]
\item \textbf{RQ1:} How does \textsc{OneLatent} compare with ID-based, SID-based, and reasoning-enhanced FRM baselines?

\item \textbf{RQ2:} Does MV-ACoT provide higher-quality and complementary CoT supervision for improving FRMs?

\item \textbf{RQ3:} Can latent reasoning further improve performance over explicit CoT reasoning?

\item \textbf{RQ4:} Does latent reasoning preserve general ability while improving reasoning efficiency?

\item \textbf{RQ5:} Can the resource-efficient latent-reasoning architecture improve online performance over a strong production baseline under the production serving architecture?
\end{itemize}

\subsection{Experimental Setting}
\subsubsection{Datasets and Models}
We evaluate \textsc{OneLatent} on both public and industrial datasets. The public dataset is drawn from the advertising domain of the Kuaishou LLM-Rec benchmark~\footnote{\url{https://huggingface.co/datasets/OpenOneRec/Explorer_LLM_Rec_Competition}}, while the industrial dataset comprises seven consecutive days of real-world user interactions collected from Kuaishou’s local-services platform. Table~\ref{tab:dataset_statistics} summarizes the basic statistics of both datasets. Unless otherwise specified, our experiments use Qwen3.5-0.8B~\cite{qwen3.5} as the backbone; we additionally evaluate a \textsc{OneLatent}-9B checkpoint on the industrial dataset. Before Stage~I, we augment the vocabulary with SID tokens and three latent tokens, after which all stages are trained with full-parameter updates. For the public benchmark, we initialize the model from the OneReason-0.8B-pretrain-competition checkpoint~\footnote{\url{https://huggingface.co/OpenOneRec/OneReason-0.8B-pretrain-competition}}~\cite{onereason2026}, and then perform latent-token alignment followed by curriculum-based multi-stage post-training.

\begin{table}[h]
\centering
\caption{Statistics of the datasets used in our experiments.}
\renewcommand{\arraystretch}{1.08}
\begin{tabular*}{\columnwidth}{@{\extracolsep{\fill}}lccc@{}}
\toprule
\textbf{Dataset} & \textbf{SFT Train} & \textbf{CoT Train} & \textbf{Test} \\
\midrule
Kuaishou Locallife & 100,063 & 59,113 & 20,393 \\
Kuaishou LLM-Rec & 60,000 & 30,000 & 10,000 \\
\bottomrule
\end{tabular*}
\label{tab:dataset_statistics}
\end{table}

\subsubsection{Baselines}
We compare \textsc{OneLatent} with three groups of baselines: (1) ID-based sequential recommenders, including GRU4Rec~\cite{hidasi2015session}, BERT4Rec~\cite{sun2019bert4rec}, and HSTU~\cite{zhai2024actions}; (2) SID-based generative recommenders, including TIGER~\cite{rajput2023recommender} and OneRec~\cite{zhou2025onerec}; and (3) reasoning-enhanced FRM baselines derived from OneReason, including the standard OneReason model and the \textit{No-Think} and \textit{Think} variants of OneReason-CoT. Think and No-Think use the exact same trained checkpoint and differ only in the control token supplied at inference. \textit{Think} generates an explicit rationale before the target SID, whereas \textit{No-Think} directly generates the target SID. The generation limit accommodates the complete rationale and final SID, and we observed no SID truncation caused by overlong rationales. For CoT analysis, we compare MV-ACoT with standard OneReason-CoT and our Adaptive CoT.

\subsubsection{Metrics and Implementation}
For recommendation performance, we report SID Hit@$K$ and ID Hit@$K$ at $K\in\{1,16,32,64\}$. For SID-based models, generated SIDs are mapped back to item IDs through SID decoding. For latent reasoning analysis, we report codebook retrieval Hit@1/2/3, inference QPS, semantic similarity, and MMLU scores. We implement \textsc{OneLatent} with the Hugging Face \texttt{Trainer} and Fully Sharded Data Parallel (FSDP) for distributed training. For evaluation, we employ vLLM with beam search for efficient decoding. More implementation details are provided in Appendix~\ref{app:parameters}.

\begin{table*}[tp]
\centering

\caption{Results on public and industrial datasets. ID-based baselines do not generate semantic IDs, and their SID results are marked as ``--''. OR/Ada/Exp./Eval./Attr. denote OneReason-CoT, Adaptive CoT, and MV-ACoT views. The final row reports the relative gain of \textsc{OneLatent}--Merge over OneReason+CoT Think--OR.}

\begingroup
\setlength{\tabcolsep}{1.5pt}
\resizebox{\textwidth}{!}{
\begin{tabular}{lllcccccccccccccccc}
\toprule
\multirow{2}{*}{\textbf{Backbone}}
& \multirow{2}{*}{\textbf{Method}}
& \multirow{2}{*}{\textbf{CoT}}
& \multicolumn{8}{c}{\textbf{OneReason Public Benchmark}}
& \multicolumn{8}{c}{\textbf{Industrial Dataset}} \\
\cmidrule(lr){4-11} \cmidrule(lr){12-19}
& & 
& \textbf{SID@1} & \textbf{SID@16} & \textbf{SID@32} & \textbf{SID@64}
& \textbf{ID@1} & \textbf{ID@16} & \textbf{ID@32} & \textbf{ID@64}
& \textbf{SID@1} & \textbf{SID@16} & \textbf{SID@32} & \textbf{SID@64}
& \textbf{ID@1} & \textbf{ID@16} & \textbf{ID@32} & \textbf{ID@64} \\
\midrule

\multirow{3}{*}{ID-based}
& GRU4Rec
& / & -- & -- & -- & -- & 0.0119 & 0.0147 & 0.0157 & 0.0182
& -- & -- & -- & -- & 0.0158 & 0.1057 & 0.1454 & 0.1883 \\

& BERT4Rec
& / & -- & -- & -- & -- & 0.0122 & 0.0153 & 0.0161 & 0.0189
& -- & -- & -- & -- & 0.0174 & 0.1074 & 0.1480 & 0.1932 \\

& HSTU
& / & -- & -- & -- & -- & 0.0114 & 0.0146 & 0.0165 & 0.0195
& -- & -- & -- & -- & 0.0145 & 0.1039 & 0.1414 & 0.1812 \\
\midrule

\multirow{2}{*}{SID-based}
& TIGER
& / & 0.0051 & 0.0061 & 0.0076 & 0.0078 & 0.0106 & 0.0124 & 0.0137 & 0.0156
& 0.0714 & 0.1000 & 0.1197 & 0.1432 & 0.0092 & 0.0401 & 0.0598 & 0.0859 \\

& OneRec
& / & 0.0049 & 0.0063 & 0.0076 & 0.0082 & 0.0108 & 0.0135 & 0.0152 & 0.0177
& 0.0708 & 0.1031 & 0.1252 & 0.1468 & 0.0116 & 0.0447 & 0.0650 & 0.0900 \\
\midrule

\multirow{1}{*}{LLM-based}
& OneReason
& / & 0.0055 & 0.0077 & 0.0079 & 0.0081 & 0.0234 & 0.0244 & 0.0258 & 0.0272
& 0.0762 & 0.1527 & 0.1603 & 0.1658 & 0.2254 & 0.4522 & 0.4750 & 0.4941 \\
\midrule

\multirow{6}{*}{LLM-based}
& \multirow{6}{*}{\makecell[c]{OneReason+CoT\\NoThink}}
& OR
& 0.0065 & 0.0103 & 0.0104 & 0.0104 & 0.0262 & 0.0306 & 0.0314 & 0.0334
& 0.0809 & 0.1639 & 0.1739 & 0.1823 & 0.2293 & 0.4499 & 0.4727 & 0.4936 \\

& & Ada
& 0.0070 & 0.0113 & 0.0124 & 0.0130 & 0.0270 & 0.0341 & 0.0364 & 0.0387
& 0.0873 & 0.1746 & 0.1832 & 0.1913 & 0.2365 & 0.4620 & 0.4850 & 0.5049 \\

& & Exp.
& 0.0068 & 0.0112 & 0.0113 & 0.0113 & 0.0228 & 0.0286 & 0.0318 & 0.0374
& 0.0881 & 0.1728 & 0.1819 & 0.1894 & 0.2385 & 0.4621 & 0.4845 & 0.5032 \\

& & Eval.
& 0.0079 & 0.0139 & 0.0172 & 0.0213 & 0.0252 & 0.0357 & 0.0486 & 0.0623
& 0.0894 & 0.1800 & 0.1896 & 0.1962 & 0.2356 & 0.4630 & 0.4881 & 0.5065 \\

& & Attr.
& 0.0071 & 0.0125 & 0.0165 & 0.0208 & 0.0255 & 0.0379 & 0.0462 & 0.0601
& 0.0897 & 0.1793 & 0.1876 & 0.1943 & 0.2392 & 0.4640 & 0.4860 & 0.5040 \\

& & Merge
& 0.0082 & 0.0131 & 0.0178 & 0.0209 & 0.0258 & 0.0362 & 0.0471 & 0.0612
& 0.0861 & 0.1720 & 0.1790 & 0.1851 & 0.2385 & 0.4640 & 0.4846 & 0.5038 \\
\midrule

\multirow{6}{*}{LLM-based}
& \multirow{6}{*}{\makecell[c]{OneReason+CoT\\Think}}
& OR
& 0.0063 & 0.0073 & 0.0080 & 0.0089 & 0.0235 & 0.0263 & 0.0278 & 0.0305
& 0.0601 & 0.1405 & 0.1563 & 0.1697 & 0.1838 & 0.4080 & 0.4450 & 0.4759 \\

& & Ada
& 0.0064 & 0.0074 & 0.0086 & 0.0093 & 0.0233 & 0.0267 & 0.0296 & 0.0311
& 0.0714 & 0.1486 & 0.1590 & 0.1688 & 0.2077 & 0.4329 & 0.4606 & 0.4844 \\

& & Exp.
& 0.0065 & 0.0082 & 0.0084 & 0.0087 & 0.0235 & 0.0285 & 0.0291 & 0.0299
& 0.0672 & 0.1190 & 0.1391 & 0.1465 & 0.2013 & 0.3992 & 0.4301 & 0.4512 \\

& & Eval.
& 0.0065 & 0.0081 & 0.0087 & 0.0096 & 0.0234 & 0.0276 & 0.0299 & 0.0328
& 0.0696 & 0.1169 & 0.1332 & 0.1487 & 0.2083 & 0.3575 & 0.3952 & 0.4303 \\

& & Attr.
& 0.0066 & 0.0078 & 0.0081 & 0.0084 & 0.0235 & 0.0273 & 0.0282 & 0.0296
& 0.0682 & 0.1287 & 0.1452 & 0.1617 & 0.2029 & 0.3739 & 0.4088 & 0.4458 \\

& & Merge
& 0.0072 & 0.0084 & 0.0086 & 0.0091 & 0.0241 & 0.0281 & 0.0288 & 0.0303
& 0.0728 & 0.1301 & 0.1400 & 0.1505 & 0.2127 & 0.3985 & 0.4275 & 0.4567 \\
\midrule

\multirow{7}{*}{LLM-based}
& \multirow{7}{*}{\textsc{OneLatent}}
& OR
& 0.0065 & 0.0082 & 0.0089 & 0.0092 & 0.0156 & 0.0253 & 0.0268 & 0.0282
& 0.0882 & 0.1771 & 0.1867 & 0.1939 & 0.2354 & 0.4632 & 0.4856 & 0.5061 \\

& & Ada
& 0.0075 & 0.0126 & 0.0138 & 0.0144 & 0.0224 & 0.0378 & 0.0403 & 0.0429
& 0.0898 & 0.1770 & 0.1869 & 0.1935 & 0.2376 & 0.4631 & 0.4867 & 0.5052 \\

& & Exp.
& 0.0072 & 0.0125 & 0.0137 & 0.0144 & 0.0179 & 0.0350 & 0.0373 & 0.0402
& 0.0930 & 0.1811 & 0.1908 & 0.1971 & 0.2425 & 0.4650 & 0.4883 & 0.5060 \\

& & Eval.
& 0.0087 & 0.0155 & 0.0189 & 0.0233 & 0.0300 & 0.0410 & 0.0560 & 0.0712
& 0.0946 & 0.1831 & 0.1931 & 0.1989 & 0.2436 & 0.4660 & 0.4891 & 0.5070 \\

& & Attr.
& 0.0072 & 0.0145 & 0.0150 & 0.0167 & 0.0299 & 0.0365 & 0.0478 & 0.0560
& 0.0902 & 0.1779 & 0.1869 & 0.1936 & 0.2396 & 0.4645 & 0.4879 & 0.5055 \\

& & \cellcolor{oursgray}\textbf{Merge}
& \cellcolor{oursgray}\textbf{0.0089}
& \cellcolor{oursgray}\textbf{0.0158}
& \cellcolor{oursgray}\textbf{0.0193}
& \cellcolor{oursgray}\textbf{0.0239}
& \cellcolor{oursgray}\textbf{0.0331}
& \cellcolor{oursgray}\textbf{0.0506}
& \cellcolor{oursgray}\textbf{0.0584}
& \cellcolor{oursgray}\textbf{0.0742}
& \cellcolor{oursgray}\textbf{0.0949}
& \cellcolor{oursgray}\textbf{0.1847}
& \cellcolor{oursgray}\textbf{0.1957}
& \cellcolor{oursgray}\textbf{0.1993}
& \cellcolor{oursgray}\textbf{0.2440}
& \cellcolor{oursgray}\textbf{0.4662}
& \cellcolor{oursgray}\textbf{0.4895}
& \cellcolor{oursgray}\textbf{0.5088} \\

& & \makecell[c]{\textit{Gain over}\\\textit{Think--OR ($\Delta$\%)}}
& \textcolor{gainred}{\textbf{+41\%$\uparrow$}}
& \textcolor{gainred}{\textbf{+116\%$\uparrow$}}
& \textcolor{gainred}{\textbf{+141\%$\uparrow$}}
& \textcolor{gainred}{\textbf{+169\%$\uparrow$}}
& \textcolor{gainred}{\textbf{+41\%$\uparrow$}}
& \textcolor{gainred}{\textbf{+92\%$\uparrow$}}
& \textcolor{gainred}{\textbf{+110\%$\uparrow$}}
& \textcolor{gainred}{\textbf{+143\%$\uparrow$}}
& \textcolor{gainred}{\textbf{+58\%$\uparrow$}}
& \textcolor{gainred}{\textbf{+31\%$\uparrow$}}
& \textcolor{gainred}{\textbf{+25\%$\uparrow$}}
& \textcolor{gainred}{\textbf{+17\%$\uparrow$}}
& \textcolor{gainred}{\textbf{+33\%$\uparrow$}}
& \textcolor{gainred}{\textbf{+14\%$\uparrow$}}
& \textcolor{gainred}{\textbf{+10\%$\uparrow$}}
& \textcolor{gainred}{\textbf{+7\%$\uparrow$}} \\
\bottomrule
\end{tabular}
}
\endgroup

\label{tab:OneLatent_benchmark_compact}
\end{table*}

\subsection{Overall Results (RQ1)}

\paragraph{Performance on Different Benchmarks.}
Table \ref{tab:OneLatent_benchmark_compact} reports the overall recommendation performance on both datasets. \textsc{OneLatent} denotes the latent-reasoning model trained with the indicated CoT construction. \textsc{OneLatent}--Merge achieves the best value among the compared methods for every reported metric. Relative to the strongest non-\textsc{OneLatent} result in each column, its gains at $K=64$ are 12.21\% and 19.10\% for public SID and ID retrieval, and 1.58\% and 0.45\% for industrial SID and ID retrieval. The final table row serves a different, explicitly labeled diagnostic comparison against the matched OneReason-CoT Think--OR configuration. It is not a claim against the strongest baseline. On the industrial dataset, the corresponding SID@64 gains over Think--OR and No-Think--OR are 17.44\% and 9.33\%, respectively.


The Think/No-Think gap cannot be attributed to initialization or training checkpoints because both modes share identical weights and are selected by control tokens. Think must nevertheless autoregressively generate a natural-language rationale before the SID, creating a longer decision path more exposed to rationale errors and text--SID mismatch. This is consistent with OneReason~\cite{team2026onereason}, where thinking does not necessarily outperform non-thinking recommendation, and the PauseRec~\cite{he2026implicit} analysis of explicit reasoning. We therefore treat Think as an explicit-decoding baseline, rather than assuming that additional generated reasoning must improve accuracy. In contrast, without complex RL training, our \textsc{OneLatent} consistently makes Think outperform No-Think, demonstrating that latent reasoning reliably improves recommendation accuracy.

\paragraph{Scalability across backbone sizes.}

To examine whether the benefit of latent reasoning transfers beyond the 0.8B setting, we additionally evaluate the \textsc{OneLatent}-9B checkpoint on the industrial dataset. Table~\ref{tab:backbone_scalability} compares direct recommendation, explicit CoT inference, and latent reasoning under matched backbone sizes. For a controlled comparison, the No-Think, Think, and \textsc{OneLatent} rows use the same MV-ACoT--Merge supervision. At the 9B scale, \textsc{OneLatent} achieves a SID@64 of 0.2008, outperforming the strongest reported non-latent 9B result of 0.1857 by 8.13\%. Notably, the 9B backbone provides stronger direct recommendation performance than the 0.8B backbone, whereas their performance becomes comparable after training with MV-ACoT supervision, demonstrating the effectiveness of the training recipe across model scales. Moreover, the consistent advantage over size-matched direct and explicit CoT baselines indicates that the benefit of latent reasoning transfers to a substantially larger FRM.

\begin{table}[t]
\centering
\caption{Backbone scalability on the industrial dataset. ``Direct'' denotes standard next-item SFT without CoT. The No-Think, Think, and \textsc{OneLatent} rows use MV-ACoT--Merge supervision.}
\label{tab:backbone_scalability}
\small
\setlength{\tabcolsep}{3.2pt}
\renewcommand{\arraystretch}{1.02}
\begin{tabular*}{\columnwidth}{@{\extracolsep{\fill}}llcccc@{}}
\toprule
\textbf{Backbone} & \textbf{Mode} & \textbf{SID@1} & \textbf{SID@16} & \textbf{SID@32} & \textbf{SID@64} \\
\midrule
0.8B & Direct   & 0.0762 & 0.1527 & 0.1603 & 0.1658 \\
& No-Think & 0.0861 & 0.1720 & 0.1790 & 0.1851 \\
& Think    & 0.0728 & 0.1301 & 0.1400 & 0.1505 \\
& \textsc{OneLatent} & \textbf{0.0949} & \textbf{0.1847} & \textbf{0.1957} & \textbf{0.1993} \\
\midrule
9B & Direct   & 0.0818 & 0.1681 & 0.1787 & 0.1857 \\
& No-Think & 0.0833 & 0.1647 & 0.1757 & 0.1825 \\
& Think    & 0.0691 & 0.1251 & 0.1351 & 0.1437 \\
& \textsc{OneLatent} & \textbf{0.0943} & \textbf{0.1843} & \textbf{0.1940} & \textbf{0.2008} \\
\bottomrule
\end{tabular*}
\end{table}

\begin{table}[t]
\centering
\caption{LLM-as-a-judge evaluation of CoT generation strategies on a 1--5 scale (higher is better). \textsc{OR} and \textsc{Ada} denote OneReason-CoT and Adaptive CoT, while \textsc{Exp.}, \textsc{Eval.}, and \textsc{Attr.} denote the three MV-ACoT views. The best result in each row is shown in bold.}
\label{tab:cot_quality_eval}
\setlength{\tabcolsep}{3.2pt}
\renewcommand{\arraystretch}{0.92}
\begin{tabular*}{\columnwidth}{@{\extracolsep{\fill}}lccccc}
\toprule
\textbf{Criterion}
& \textsc{OR}
& \textsc{Ada}
& \textsc{Exp.}
& \textsc{Eval.}
& \textsc{Attr.} \\
\midrule
Factuality               & 3.48 & 3.70 & 3.96 & \textbf{4.04} & 3.62 \\
Evidence Use             & 3.17 & 3.68 & 4.00 & \textbf{4.34} & 4.13 \\
Signal Strength          & 3.19 & 3.35 & 3.74 & \textbf{4.04} & 3.52 \\
Intent Translation       & 3.76 & 3.86 & 4.52 & 4.59 & \textbf{4.68} \\
Causal Logic             & 2.74 & 3.27 & 4.02 & \textbf{4.25} & 3.67 \\
Ad--User Fit             & 3.26 & 3.85 & 4.49 & \textbf{4.58} & 4.30 \\
Spatiotemporal Reasoning & 3.34 & 3.87 & 4.32 & 4.44 & \textbf{4.64} \\
Confidence               & 2.78 & 2.91 & 3.91 & \textbf{4.11} & 3.64 \\
Task Fulfillment         & 2.56 & 3.36 & 4.48 & \textbf{4.59} & 4.08 \\
\midrule
\textbf{Mean}            & 3.14 & 3.54 & 4.16 & \textbf{4.33} & 4.03 \\
\bottomrule
\end{tabular*}
\end{table}

\subsection{CoT Analysis (RQ2)}

We analyze MV-ACoT from three aspects: teacher-CoT quality, adaptive reasoning budget, and recommendation performance.

\subsubsection{LLM-as-a-Judge Evaluation}
We evaluate generated reasoning traces using an LLM-as-a-Judge protocol. For each CoT construction view, we sample 100 instances and score them with a unified rubric covering nine dimensions: \emph{factuality}, \emph{evidence selection}, \emph{signal strength}, \emph{intent translation}, \emph{causal logic}, \emph{ad--user fit}, \emph{spatiotemporal reasoning}, \emph{confidence}, and \emph{task fulfillment}. We use GPT-5.5 as the judge; the complete prompt and rubric are provided in Appendix~\ref{app:cot_quality_prompt}.

As shown in Table \ref{tab:cot_quality_eval}, the results support the value of multi-view construction. \textsc{Exploration} produces diverse intent hypotheses that target-conditioned tasks may overlook. \textsc{Evaluation} measures how well the observed positive item matches the available evidence and outputs a 1--5 score in addition to its rationale. \textsc{Attribution} explains the evidence associated with a known positive interaction. Together, the views provide open-ended intent discovery, positive-item matching, and outcome attribution rather than repeating a single reasoning objective.

\subsubsection{Adaptive Reasoning Budget}
We quantify the effect of the prompt-guided adaptive policy using generated CoT length. Setting the average token count of the non-adaptive prompt to 100, Adaptive CoT reduces the normalized average to 65, a 35\% reduction. To verify that this reduction does not merely collapse difficult traces, we define complex cases as the top decile by non-adaptive CoT length and compare the paired adaptive and non-adaptive outputs with Qwen3-Emb-8B. Their mean semantic similarity is approximately 0.92. This result measures semantic preservation rather than task accuracy, but indicates that the prompt shortens routine reasoning while retaining most content on long-reasoning cases.

\subsubsection{Ablation Study on CoT Effectiveness}

\begin{table}[t]

\centering
\caption{Ablation of CoT construction in \textsc{OneLatent}. The final row reports the relative gain of MV-ACoT--Merge over the strongest single-view configuration in each column.}
\vspace{-2mm}

\setlength{\tabcolsep}{3.6pt}
\renewcommand{\arraystretch}{0.92}
\begin{tabular*}{\columnwidth}{@{\extracolsep{\fill}}lcccc}
\toprule
\multirow{2}{*}{\textbf{CoT Construction}}
& \multicolumn{2}{c}{\textbf{Public Benchmark}}
& \multicolumn{2}{c}{\textbf{Industrial Dataset}} \\
\cmidrule(lr){2-3}
\cmidrule(l){4-5}
& \textbf{SID@64} & \textbf{ID@64}
& \textbf{SID@64} & \textbf{ID@64} \\
\midrule
\textsc{OneReason-CoT}
& 0.0092 & 0.0282
& 0.1939 & 0.5061 \\

\textsc{Adaptive CoT}
& 0.0144 & 0.0429
& 0.1935 & 0.5052 \\

\textsc{Exploration}
& 0.0144 & 0.0402
& 0.1971 & 0.5060 \\

\textsc{Evaluation}
& 0.0233 & 0.0712
& 0.1989 & 0.5070 \\

\textsc{Attribution}
& 0.0167 & 0.0560
& 0.1936 & 0.5055 \\

\rowcolor{oursgray}
\textsc{MV-ACoT--Merge}
& \textbf{0.0239} & \textbf{0.0742}
& \textbf{0.1993} & \textbf{0.5088} \\
\midrule
\rowcolor{deltagray}
\multicolumn{1}{r}{\textit{Gain over best view}}
& \textcolor{gainred}{\textbf{+2.58\%$\uparrow$}}
& \textcolor{gainred}{\textbf{+4.21\%$\uparrow$}}
& \textcolor{gainred}{\textbf{+0.20\%$\uparrow$}}
& \textcolor{gainred}{\textbf{+0.36\%$\uparrow$}} \\
\bottomrule
\end{tabular*}
\label{tab:cot_ablation}

\end{table}

We further examine whether MV-ACoT supervision translates into downstream gains for \textsc{OneLatent}. As shown in Table~\ref{tab:cot_ablation}, Adaptive CoT substantially improves the public-benchmark metrics over OneReason-CoT while remaining comparable on the industrial dataset. Together with the 35\% reduction in average reasoning length, this shows that prompt-guided adaptation can reduce redundant supervision without uniformly degrading downstream performance.

\begin{figure}[h]
\centering
\includegraphics[width=\columnwidth]{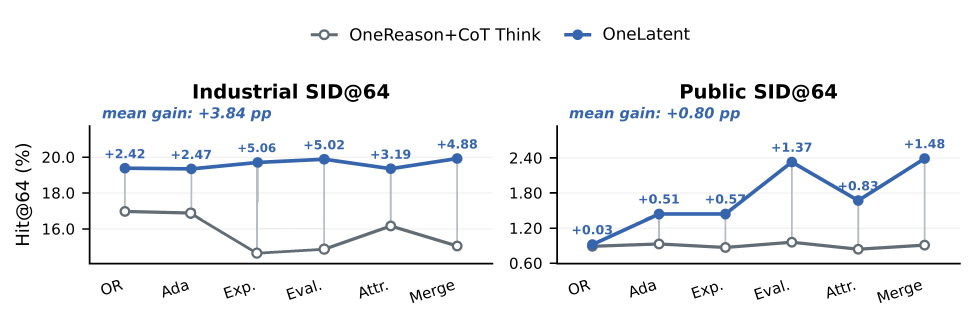}

\caption{SID@64 comparison between OneReason-Think and \textsc{OneLatent} under different CoT strategies.}
\Description{Across OR, Adaptive CoT, \textsc{Exploration}, \textsc{Evaluation}, \textsc{Attribution}, and Merge supervision, OneLatent achieves higher public and industrial SID at 64 than the matched OneReason Think mode.}

\label{fig:sid_trend_mixed_vs_latent}
\end{figure}

Single-view MV-ACoT variants provide task-specific reasoning objectives. \textsc{Evaluation} performs strongly because positive-item matching directly aligns with user--item decisions. \textsc{Exploration} broadens potential intent hypotheses, whereas \textsc{Attribution} captures conversion-oriented evidence chains. Their relative performance varies by dataset and metric, supporting their complementarity rather than implying that one view is universally optimal.
The integrated \textsc{MV-ACoT-Merge} setting further demonstrates the benefit of multi-view supervision. By combining intent exploration, evidence-grounded evaluation, and conversion attribution, it offers more comprehensive supervision than any single view. Figure~\ref{fig:sid_trend_mixed_vs_latent} also shows that \textsc{OneLatent} consistently outperforms OneReason+CoT Think under the same CoT construction strategies, suggesting that latent reasoning more effectively translates view-specific supervision into SID retrieval gains. Improvements across Adaptive CoT and MV-ACoT variants further indicate that complexity adaptation and multi-view objectives complement latent encoding.

\subsection{Latent Reasoning Analysis (RQ3)}
In this section, we examine whether latent reasoning can further improve over explicit CoT reasoning.

\subsubsection{Recommendation Performance}
We first compare \textsc{UnCoT}, \textsc{CoT}, and \textsc{OneLatent} on the recommendation task. As shown in Table~\ref{tab:latent_reasoning_effectiveness}, explicit CoT yields limited and inconsistent gains over direct prediction, whereas \textsc{OneLatent} improves performance across all SID levels. In particular, it raises Hit@L1 from $0.354$ to $0.401$ relative to explicit CoT and achieves the best Hit@L2 and Hit@L3. These results establish the effectiveness of the complete latent-token-conditioned training pipeline, although they do not by themselves isolate whether the gain is stored exclusively in the latent-token positions or distributed across the fully updated backbone.

\begin{table}[h]
\centering
\caption{Effectiveness of latent reasoning on SID prediction.}
\renewcommand{\arraystretch}{1.05}
\begin{tabular*}{\columnwidth}{@{\extracolsep{\fill}}lccc@{}}
\toprule
\textbf{Method} & \textbf{Hit@L1} & \textbf{Hit@L2} & \textbf{Hit@L3} \\
\midrule
\textsc{OneReason-UnCoT} & 0.350 & 0.122 & 0.045 \\
\textsc{OneReason-CoT} & 0.354 & 0.124 & 0.043 \\
\textsc{OneLatent}
& \textbf{0.401} & \textbf{0.143} & \textbf{0.071} \\
\bottomrule
\end{tabular*}

\label{tab:latent_reasoning_effectiveness}
\end{table}

\subsubsection{Inference Efficiency}
We compare the inference efficiency of \textsc{UnCoT}, \textsc{CoT}, and \textsc{OneLatent} on 872 test instances. As shown in Table~\ref{tab:inference_efficiency}, explicit CoT must autoregressively generate a rationale before the SID, substantially increasing latency and reducing throughput. In contrast, \textsc{OneLatent} decodes only the SID after three fixed latent token IDs, achieving latency and throughput nearly identical to \textsc{UnCoT} and a $17.49\times$ QPS gain over explicit CoT.

\begin{table}[h]
\centering
\caption{Inference efficiency comparison. Speedup is measured relative to explicit CoT.}

\renewcommand{\arraystretch}{1.05}
\begin{tabular*}{\columnwidth}{@{\extracolsep{\fill}}lccc@{}}
\toprule
\textbf{Method} & \textbf{Time (s)} $\downarrow$ & \textbf{QPS} $\uparrow$ & \textbf{Speedup} $\uparrow$ \\
\midrule
\textsc{OneReason-UnCoT} & {3.17} & {274.56} & $17.51\times$ \\
\textsc{OneReason-CoT} & 55.59 & 15.68 & $1.00\times$ \\
\textsc{OneLatent} & 3.17 & 274.29 & \textbf{17.49$\times$} \\
\bottomrule
\end{tabular*}

\label{tab:inference_efficiency}
\end{table}

\subsubsection{Same-Checkpoint Latent-Token Interventions}
\label{sec:latent-token-interventions}
To isolate whether predictions from the final model depend on the learned latent token interface, we intervene only on its inference prefix. All variants in Table~\ref{tab:latent_token_intervention} use the same final \textsc{OneLatent}-0.8B checkpoint, industrial test examples, and decoding configuration; no variant is retrained. Starting from the learned order $[\langle L_1\rangle;\langle L_2\rangle;\langle L_3\rangle]$, we (i) remove one token at a time, (ii) permute the three learned tokens, or (iii) replace them with three independently sampled sets of ordinary vocabulary tokens. The three random-replacement rows report the individual runs rather than an unreported average.

\begin{table}[t]
\centering
\caption{Same-checkpoint interventions on the latent token prefix on the industrial dataset. Random-1/2/3 denote independently sampled ordinary-token replacement sets. The final row reports the relative gain of the intact prefix over the strongest intervened result in each column.}
\label{tab:latent_token_intervention}
\small
\setlength{\tabcolsep}{3.0pt}
\renewcommand{\arraystretch}{1.02}
\begin{tabular*}{\columnwidth}{@{\extracolsep{\fill}}lcccc@{}}
\toprule
\textbf{Inference prefix} & \textbf{SID@32} & \textbf{SID@64} & \textbf{ID@32} & \textbf{ID@64} \\
\midrule
\rowcolor{oursgray}
Full $[L_1,L_2,L_3]$ & \textbf{0.1957} & \textbf{0.1993} & \textbf{0.4895} & \textbf{0.5088} \\
w/o $L_1$ & 0.1870 & 0.1936 & 0.4881 & 0.5052 \\
w/o $L_2$ & 0.1885 & 0.1956 & 0.4883 & 0.5062 \\
w/o $L_3$ & 0.1895 & 0.1960 & 0.4891 & 0.5063 \\
$[L_3,L_2,L_1]$ & 0.1891 & 0.1957 & 0.4885 & 0.5061 \\
$[L_3,L_1,L_2]$ & 0.1887 & 0.1953 & 0.4882 & 0.5058 \\
$[L_2,L_1,L_3]$ & 0.1890 & 0.1955 & 0.4882 & 0.5060 \\
Random-1 & 0.1880 & 0.1950 & 0.4877 & 0.5056 \\
Random-2 & 0.1871 & 0.1942 & 0.4879 & 0.5056 \\
Random-3 & 0.1877 & 0.1946 & 0.4875 & 0.5052 \\
\midrule
\rowcolor{deltagray}
\makecell[r]{\textit{Gain over best}\\\textit{variant ($\Delta\%$)}}
& \textcolor{gainred}{\textbf{+3.27\%$\uparrow$}}
& \textcolor{gainred}{\textbf{+1.68\%$\uparrow$}}
& \textcolor{gainred}{\textbf{+0.08\%$\uparrow$}}
& \textcolor{gainred}{\textbf{+0.49\%$\uparrow$}} \\
\bottomrule
\end{tabular*}

\end{table}

The intact learned sequence is best on all four reported metrics. Deletion consistently reduces SID accuracy, showing that no single token is entirely redundant at the larger retrieval cutoffs. Permutations preserve token identities and prefix length but remain below the learned order, while all three ordinary-token replacements also degrade performance. These controlled interventions provide direct evidence that the final checkpoint is sensitive to the identities and order of the trained latent tokens, rather than benefiting only from an arbitrary three-token prefix. They establish causal dependence on the interface under these interventions, but do not imply that all reasoning information is stored exclusively at the three positions because standard causal attention still exposes the original context.

\subsubsection{Ablation of Latent-Token Alignment Configurations}
\label{sec:alignment-stage-ablation}
We next vary the latent token alignment curriculum while keeping the downstream recommendation post-training and evaluation protocol fixed. Table~\ref{tab:alignment_stage_ablation} compares a single-view Stage~I warm-up, Stages~I--II without recommendation-oriented context alignment, and the complete Stage~I--III pipeline. The full alignment pipeline achieves the strongest result at every reported SID and ID cutoff. Among the single-view Stage~I variants, Evaluation is consistently strongest and approaches the full model on ID retrieval, whereas Attribution alone produces the largest degradation. Stages~I--II also remain below the full model; adding Stage~III raises SID@64 from 0.1962 to 0.1993 and ID@64 from 0.5057 to 0.5088. Together, these results support the complementarity of rationale alignment and recommendation-oriented context alignment. They also show that a single privileged-target Evaluation view is useful but does not fully replace the complete multi-view, multi-stage recipe.

\begin{table}[t]
\centering
\caption{Ablation of latent token alignment configurations on the industrial dataset. All variants use the same downstream recommendation post-training. The final row reports the relative gain from adding Stage~III to Stages~I--II.}
\label{tab:alignment_stage_ablation}
\small
\setlength{\tabcolsep}{3.0pt}
\renewcommand{\arraystretch}{1.02}
\begin{tabular*}{\columnwidth}{@{\extracolsep{\fill}}lcccc@{}}
\toprule
\makecell[l]{\textbf{Alignment}\\\textbf{configuration}} & \textbf{SID@32} & \textbf{SID@64} & \textbf{ID@32} & \textbf{ID@64} \\
\midrule
\rowcolor{oursgray}
Full: Stages I--III & \textbf{0.1957} & \textbf{0.1993} & \textbf{0.4895} & \textbf{0.5088} \\
Stage I (\textsc{Exploration}) & 0.1908 & 0.1971 & 0.4883 & 0.5060 \\
Stage I (\textsc{Evaluation}) & 0.1931 & 0.1989 & 0.4891 & 0.5070 \\
Stage I (\textsc{Attribution}) & 0.1869 & 0.1936 & 0.4879 & 0.5055 \\
Stages I--II & 0.1905 & 0.1962 & 0.4876 & 0.5057 \\
\midrule
\rowcolor{deltagray}
\makecell[r]{\textit{Gain over}\\\textit{Stages I--II ($\Delta\%$)}}
& \textcolor{gainred}{\textbf{+2.73\%$\uparrow$}}
& \textcolor{gainred}{\textbf{+1.58\%$\uparrow$}}
& \textcolor{gainred}{\textbf{+0.39\%$\uparrow$}}
& \textcolor{gainred}{\textbf{+0.61\%$\uparrow$}} \\
\bottomrule
\end{tabular*}

\end{table}

\subsubsection{Interpretability}
\label{sec:latent-interpretability}
We use the trained \textsc{OneLatent} checkpoint itself to generate a textual reconstruction. For the same randomly sampled 2,000 examples, we construct one input per view in the form $[P_u;\langle v\rangle;\mathbf{L}]$, deliberately omitting the interaction history. The checkpoint generates a view-specific rationale, which is paired with the teacher CoT from the same example. Qwen3-Emb-8B~\footnote{\url{https://huggingface.co/Qwen/Qwen3-Embedding-8B}} encodes each rationale into one vector, cosine similarity is computed for each pair, and the values are averaged within each view.

\begin{table}[h]
\centering
\caption{Mean cosine similarity between view-conditioned reconstructions and paired teacher CoT.}

\renewcommand{\arraystretch}{1.12}
\begin{tabular*}{\columnwidth}{@{\extracolsep{\fill}}lccc@{}}
\toprule
\textbf{Metric} & \textsc{Exploration} & \textsc{Evaluation} & \textsc{Attribution} \\
\midrule
Cosine Similarity & 0.8027 & \textbf{0.8043} & 0.7878 \\
\bottomrule
\end{tabular*}
\label{tab:semantic_similarity}
\end{table}

As shown in Table~\ref{tab:semantic_similarity}, the generated reconstructions achieve similarities of $0.788$--$0.804$ across \textsc{Exploration}, \textsc{Evaluation}, and \textsc{Attribution}. This provides evidence that generation conditioned on the profile, view token, and latent token IDs is semantically aligned with the corresponding teacher supervision. Because the model is fully fine-tuned and the profile remains visible, cosine similarity is not a measure of step-by-step faithfulness and does not prove that the latent IDs form a strict information bottleneck. We therefore use this experiment as an association and interpretability analysis, complementary to recommendation performance and inference efficiency. A qualitative case is provided in Appendix~\ref{sec:qualitative_reconstruction}.

\subsubsection{Number of Latent Tokens}
\begin{figure}[h]

\centering
\includegraphics[width=\columnwidth]{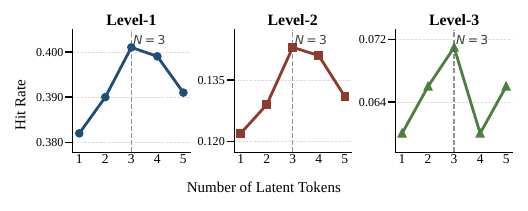}

\caption{Ablation study on the number of latent tokens.}
\Description{Hit rates at all three semantic-ID levels rise from one to three latent tokens and do not improve consistently with four or five tokens.}

\label{fig:latent_token_num_ablation}
\end{figure}

We analyze the effect of the number of latent tokens on recommendation performance. As shown in Figure~\ref{fig:latent_token_num_ablation}, Hit improves as the number of tokens increases from one to three, indicating that multiple tokens help represent fine-grained reasoning information. Further increasing the number yields no consistent gains, likely due to redundancy. We therefore use three latent tokens to balance reasoning capacity and compactness.

\subsection{General Capability Analysis (RQ4)}
\label{app:general_capability}

\begin{figure}[t]
\centering
\includegraphics[width=\columnwidth]{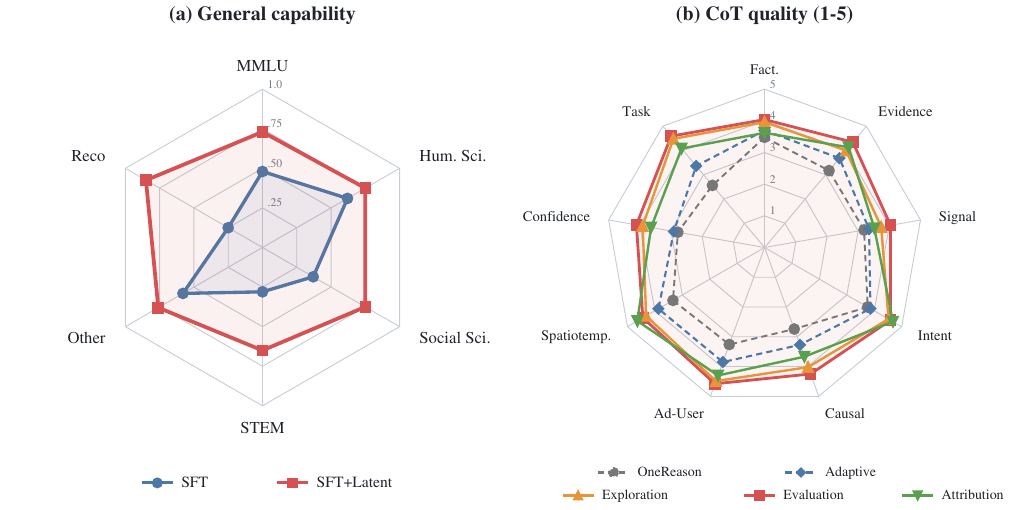}
\caption{Capability profiles (higher is better). (a) General and recommendation capability under SFT and latent supervision. (b) LLM-judge ratings (1--5) for two CoT baselines and the three MV-ACoT views.}
\Description{Paired radar charts summarize general and recommendation capability and nine dimensions of CoT quality. Evaluation leads seven judge dimensions, while Attribution leads intent translation and spatiotemporal reasoning.}
\label{fig:general_cot_radar}
\end{figure}

We further investigate whether latent reasoning preserves the model's general-purpose capabilities while enhancing recommendation performance. Figure~\ref{fig:general_cot_radar} jointly summarizes general capability and CoT quality. Panel~(a) compares models trained with explicit CoT supervision and latent-reasoning supervision across MMLU categories and recommendation quality. Our \textsc{OneLatent} consistently achieves higher scores across all evaluated MMLU categories while also improving recommendation performance, suggesting that latent supervision better preserves general knowledge and reasoning abilities while strengthening recommendation-specific reasoning. Panel~(b) plots the nine judge dimensions reported in Table~\ref{tab:cot_quality_eval}, revealing complementary strengths across the MV-ACoT views: \textsc{Evaluation} scores highest on seven dimensions, including evidence use, causal logic, and confidence, whereas \textsc{Attribution} scores highest on intent translation and spatiotemporal reasoning. \textsc{Exploration} and \textsc{Evaluation} exceed both CoT baselines on all nine dimensions; \textsc{Attribution} exceeds both on eight dimensions, with factuality below Adaptive CoT. These profiles complement the CoT-construction ablation in Table~\ref{tab:cot_ablation}.

\subsection{Online A/B Testing (RQ5)}

We conducted a seven-day online A/B test from September 3 to September 10, 2026, in a subscenario of the local-services advertising scenario in the Kuaishou App. The treatment and control groups each received 5\% of production traffic. The control group remained on the strong production online system, while the treatment group deployed \textsc{OneLatent} through the production serving architecture described in Section~\ref{sec:production-serving}. Recent user behaviors triggered value-aware and priority-scheduled model refreshes, while the latency-critical online path consumed the resulting cached candidates instead of synchronously invoking the Foundation Model.

Table~\ref{tab:online_ab} reports the results of the online A/B experiment. The treatment increases total impressions by 2.566\% and total advertising spend by 5.157\%. In addition to these directly observed metrics, the global posterior estimator reports an estimated 9.6\% lift in revenue. By combining latent reasoning with the proposed online serving architecture, the system has since been rolled out to all production traffic while avoiding the computational cost of synchronous full-traffic FRM inference. These consistent improvements demonstrate that \textsc{OneLatent} can translate its offline recommendation gains into measurable benefits under real-world traffic and production resource constraints.

\begin{table}[h]
\centering

\caption{Online A/B test results in a local-services advertising scenario in the Kuaishou app.}

\label{tab:online_ab}
\renewcommand{\arraystretch}{1.08}
\setlength{\tabcolsep}{3.5pt}
\begin{tabular*}{\columnwidth}{@{\extracolsep{\fill}}ccccc@{}}
\toprule
\textbf{Traffic} & \textbf{Days} & \textbf{Impressions} & \textbf{Ad Spend} & \textbf{Est. Revenue} \\
\midrule
$5\% $ & $7 $ & $+2.566\%$ & $+5.157\%$ & $+9.6\%$ \\
\bottomrule
\end{tabular*}
\end{table}


\section{Conclusion}
In this work, we present \textsc{OneLatent}, investigating whether FRMs can retain reasoning benefits without generating rationales during inference. We show that explicit CoT can serve as training supervision: MV-ACoT provides complementary, complexity-aware rationales, while progressive alignment distills them into a compact latent interface activated through curriculum training. Across the industrial Kuaishou dataset and the public Kuaishou LLM-Rec benchmark, this design outperforms explicit CoT approaches and conventional baselines. On the industrial dataset, \textsc{OneLatent} raises SID@64 by 17.44\% and 9.33\% relative to matched \textit{Think} and \textit{No-Think} variants, respectively, while providing more than $10\times$ higher inference throughput. Further analysis indicates that this specialization strengthens recommendation reasoning without eroding general capabilities. Our results extend beyond offline evaluation. The production architecture uses behavior-triggered scheduling and asynchronous materialization to keep FRM computation outside the critical online request path, after which generated candidates enter the existing retrieval pipeline and compete in its fine-ranking stage. In Kuaishou's local-services advertising scenario, this deployment yields an estimated 9.6\% revenue lift over a strong baseline. Taken together, these results establish \textsc{OneLatent} as an effective and deployable approach to efficient reasoning, marking an important step toward the large-scale industrial deployment of recommendation foundation models.


\balance
\bibliographystyle{ACM-Reference-Format}
\bibliography{ref}

\clearpage
\appendix
\input{appendix}

\end{document}

%% file: appendix.tex
\setlength{\emergencystretch}{1.5em}

\section{Dataset Construction and Parameter Settings}
\label{app:parameters}

\subsection{Implementation Details}
We perform full-parameter supervised fine-tuning on standard recommendation instances and CoT-augmented instances. All stages use the standard causal attention mask; no special attention bottleneck, auxiliary encoder, projector, or rationale decoder is added. Training is distributed over seven nodes with eight GPUs per node. We use a per-device batch size of 4 and one gradient-accumulation step, resulting in an effective global batch size of 224. The model is optimized using AdamW with a peak learning rate of \(1.0\times10^{-4}\), a cosine learning-rate schedule, a warmup ratio of \(0.01\), and a weight decay of \(0.01\). We train the model for two effective epochs in BF16 precision and enable gradient checkpointing to reduce memory consumption. The maximum training sequence length is 4,096 tokens. During evaluation, the Think generation budget is sufficient to contain both the complete rationale and the final SID; no SID truncation due to rationale length was observed. For the streaming dataset, the number of optimization steps is dynamically determined from the total number of samples in the two training streams:

\begin{equation}
N_{\mathrm{step}}
=
\left\lceil
\frac{
    (N_{1}+N_{2})E
}{
    N_{\mathrm{node}}
    N_{\mathrm{GPU}}
    B_{\mathrm{device}}
    N_{\mathrm{acc}}
}
\right\rceil,
\end{equation}
where \(N_{1}\) and \(N_{2}\) denote the numbers of samples in the two streams, \(E=2\) is the target number of epochs, \(N_{\mathrm{node}}=7\), \(N_{\mathrm{GPU}}=8\), \(B_{\mathrm{device}}=4\), and \(N_{\mathrm{acc}}=1\). We use a fixed random seed of 42 for all experiments.

\subsection{Supervised Dataset Construction}
We construct the supervised fine-tuning dataset from temporally ordered user-item interaction logs. Each item is associated with a hierarchical semantic identifier (SID), while each user is associated with a compact profile containing general preferences, local-life demands, and, when available, explicitly expressed intents.

For each user, we first sort all interactions chronologically and remove redundant interactions with the same item. When multiple interaction signals are associated with a duplicated item, we retain the strongest feedback according to the following priority:
\[
\text{conversion} >
\text{click} >
\text{long view} >
\text{impression}.
\]
Each conversion event is subsequently treated as a prediction target. Specifically, for the \(k\)-th conversion event, all interactions occurring before it form the historical context, whereas the SID of the converted item serves as the supervision signal. In this way, a user with multiple conversion events may contribute multiple training instances without including the target item in its own input history.

To control the context length and emphasize recent interests, we retain at most the latest 50 interactions before each target event. Historical items are represented by their hierarchical SIDs and augmented with their corresponding behavior types, allowing the model to distinguish weak feedback, such as impressions, from stronger preference signals, such as clicks and conversions. The resulting input combines the compact user profile, local-life demand, explicit intent, and behavior-aware SID sequence. The output is the hierarchical SID of the next converted item. Formally, each training instance is written as
\begin{equation}
\mathcal{H}_{u,k}
=
\bigl[(s_i, a_i)\bigr]_{i=\max(1,\,k-L)}^{k-1},
\end{equation}

\begin{equation}
\mathcal{X}_{u,k}
=
[P_u;\, D_u;\, I_u;\, \mathcal{H}_{u,k}],
\qquad
\mathcal{Y}_{u,k} = s_k,
\end{equation}
where \(P_u\), \(D_u\), and \(I_u\) denote the user profile, user demand, and explicit intent, respectively; \(s_i\) and \(a_i\) denote the SID and behavior type of the \(i\)-th interaction; and \(L=50\) is the maximum history length.
We discard instances whose user profile or target SID is unavailable. Historical items without valid SID mappings are filtered from the input sequence. Finally, all valid samples are stored in sharded JSONL files to support memory-efficient streaming and distributed training.

\begin{table}[!h]
\centering
\caption{Hyperparameters for multi-stream SFT.}
\label{tab:multi_stream_hyperparameters}
\renewcommand{\arraystretch}{1.08}
\setlength{\tabcolsep}{3pt}
\begin{tabular}{
@{}
>{\RaggedRight\arraybackslash}p{0.21\linewidth}
>{\RaggedRight\arraybackslash}p{0.43\linewidth}
>{\RaggedRight\arraybackslash}p{0.30\linewidth}
@{}
}
\toprule
\textbf{Category} & \textbf{Hyperparameter} & \textbf{Value} \\
\midrule

\multirow{5}{*}{\makecell[l]{Model\\configuration}}
& Initialization & Qwen3.5-0.8B \\
& Fine-tuning strategy & Full-parameter fine-tuning \\
& Numerical precision & BF16 \\
& Maximum sequence length & 4,096 tokens \\
& Number of latent token IDs & 3 \\
\midrule

\multirow{2}{*}{\makecell[l]{Semantic\\IDs}}
& Number of semantic levels & 3 \\
& Codebook size per level & 1,024 \\
\midrule

\multirow{6}{*}{Optimization}
& Optimizer & AdamW \\
& Learning rate & $1.0 \times 10^{-4}$ \\
& Scheduler & Cosine \\
& Warmup ratio & 0.01 \\
& Weight decay & 0.01 \\
& Gradient checkpointing & Enabled \\
\midrule

\multirow{3}{*}{Training}
& Epochs & 2 \\
& Gradient accumulation steps & 1 \\
& Random seed & 42 \\
\midrule

\multirow{2}{*}{\makecell[l]{Streaming\\data}}
& Number of training streams & 2 \\
& Maximum training steps & $\left\lceil 2(N_1 + N_2) / 224 \right\rceil$ \\
\midrule

\multirow{3}{*}{Checkpoint}
& Logging interval & 10 steps \\
& Checkpoint interval & 500 steps \\
& Maximum retained checkpoints & 5 \\
\midrule

\multirow{2}{*}{Hardware}
& Number of nodes & 7 \\
& GPUs per node & 8 \\
\bottomrule
\end{tabular}
\end{table}

\subsection{CoT Supervision Construction}
We employ Qwen3-235B-A22B-Instruct-2507\footnote{\url{https://huggingface.co/Qwen/Qwen3-235B-A22B-Instruct-2507}} as the teacher model. Exploration uses the user profile and historical interactions and generates a textual rationale plus multiple potential intents, without observing a candidate or target. Evaluation uses the observed positive item as privileged teacher-side guidance and generates a textual rationale plus a 1--5 matching score; no negative candidates are sampled. Attribution is explicitly told that the observed item is positive and generates only a textual rationale explaining the supporting evidence. The same teacher-side positive item may thus shape Evaluation and Attribution targets, but neither its description nor its SID is included in the Stage I/II student prefix.

\begin{table}[h]
\centering
\caption{Exact student prefixes and outputs. Candidate descriptions, positive-item descriptions, and target SIDs never appear in a student prefix.}
\label{tab:student_sequence_formats}
\small
\setlength{\tabcolsep}{3pt}
\renewcommand{\arraystretch}{1.08}
\begin{tabular}{@{}
>{\RaggedRight\arraybackslash}p{0.16\linewidth}
>{\RaggedRight\arraybackslash}p{0.36\linewidth}
>{\RaggedRight\arraybackslash}p{0.40\linewidth}@{}}
\toprule
\textbf{Phase} & \textbf{Student prefix} & \textbf{Output} \\
\midrule
Stages I--II & $[P_u;H_u;\langle v\rangle;\mathbf{L}]$ & View-specific teacher text, including intent list or 1--5 score when applicable \\
Stage III & $[P_u;H_u;\mathcal{C}_u;\mathbf{L}]$ & Positive target SID \\
Direct post-training & $[P_u;H_u;\mathcal{C}_u]$ & Positive target SID \\
Latent post-training & $[P_u;H_u;\mathcal{C}_u;\mathbf{L}]$ & Positive target SID \\
Online inference & $[P_u;H_u;\mathcal{C}_u;\mathbf{L}]$ & Generated target SID \\
\bottomrule
\end{tabular}
\end{table}

\begin{table}[h]
\centering
\caption{Configuration for constructing CoT supervision.}
\label{tab:cot_construction}

\renewcommand{\arraystretch}{1.08}
\setlength{\tabcolsep}{3pt}

\begin{tabular}{
@{}
>{\RaggedRight\arraybackslash}p{0.25\linewidth}
>{\RaggedRight\arraybackslash}p{0.67\linewidth}
@{}
}
\toprule
\textbf{Component} & \textbf{Configuration} \\
\midrule

Teacher model
& Qwen3-235B-A22B-Instruct-2507 \\
\midrule

User information
& User profile, user demand, and explicitly expressed intent \\
\midrule

Interaction history
& Item descriptions, behavior types, timestamps, and geographic context \\
\midrule

Exploration output
& Textual rationale and multiple potential intents; no candidate or target is provided to the teacher \\
\midrule

Evaluation output
& Textual rationale and a 1--5 match score, generated from the real positive item only \\
\midrule

Attribution output
& Textual rationale for an item explicitly identified to the teacher as a positive interaction \\
\midrule

Student-side exclusion
& No candidate description, positive-item description, target SID, or negative candidate in the Stage I/II prefix \\
\bottomrule
\end{tabular}
\end{table}

\section{Latent Reconstruction Case Study}
\label{sec:qualitative_reconstruction}

A qualitative comparison between the generated reconstruction and explicit CoT shows that the two texts share the main commercial semantics, including men's-health interest, potential lead-generation intent, price sensitivity, offline fulfillment, and dense L3 behavioral signals. The reconstruction is generated by the trained \textsc{OneLatent} checkpoint from $[P_u;\langle v\rangle;\mathbf{L}]$, without the interaction history and without a separate decoder. The two traces are largely aligned on price sensitivity, consultation intent, and conversion-funnel stage, while differing in their assessment of geographic accessibility.

With an embedding similarity of approximately $0.80$, this case illustrates semantic alignment without requiring identical wording. Because the user profile remains visible and the whole checkpoint is fine-tuned, it should not be interpreted as proof that the latent token IDs alone store the rationale or form a strict information bottleneck.

\section{MV-ACoT Prompt Design}

\label{app:mv_acot_prompt}
Beyond the complementary Exploration, Evaluation, and Attribution tasks, MV-ACoT introduces business-aware prompt constraints to improve the quality and reliability of teacher-generated reasoning. These constraints ground the teacher model in local-service lead-ad recommendation, requiring it to translate content interests into actionable commercial demands and prioritize strong behavioral evidence over weak exposures. This prevents reliance on generic content preferences that are insufficient for conversion-oriented decisions.

MV-ACoT further adapts reasoning paths to the strength, consistency, and uncertainty of user signals. The prompt first asks the teacher to assess complexity from the available strong behaviors and conflicting or noisy signals. It then selects only relevant dimensions, such as fulfillment mode, geography, time, industry constraints, and conversion stage. Clear cases should converge quickly and summarize secondary noise, whereas complex cases may compare multiple plausible explanations. This is an in-prompt self-assessment: no external difficulty classifier, binary label, threshold, or reported classification accuracy is involved.

\input{appendix_eval_tables}

\section{CoT Quality Evaluation Prompt}
\label{app:cot_quality_prompt}

We employ an LLM-as-a-Judge protocol to evaluate the quality of teacher-generated reasoning traces for local-service lead-ad recommendation. The protocol evaluates whether a trace derives a calibrated judgment from available user evidence while respecting the objective of its view. For each instance, the judge receives the task type, history, user profile, the positive item for Evaluation or Attribution, and the generated trace.

User history is treated as the primary factual evidence. User-profile attributes and target information may provide context, but must not be backfilled into historical behaviors or used to fabricate supporting evidence. As summarized in Tables~\ref{tab:judge_task_principles} and~\ref{tab:judge_dimensions}, the protocol combines task-specific requirements with a unified multi-dimensional scoring rubric. This design evaluates not only factuality and evidence selection, but also intent translation, causal logic, advertising fit, spatiotemporal reasoning, confidence calibration, and task fulfillment.

\noindent\colorbox{black!7}{%
\begin{minipage}{\dimexpr\linewidth-2\fboxsep\relax}

\scriptsize
\textbf{Evaluation Instance Format.}

\begin{itemize}[leftmargin=1.25em,itemsep=1pt,topsep=2pt]
    \item \textbf{Task type:} \texttt{\{\{task\_type\}\}}
    
    \item \textbf{Historical behaviors:} \texttt{\{\{history\_text\}\}}
    
    \item \textbf{User profile:} \texttt{\{\{user\_plain\_text\}\}}
    
    \item \textbf{Positive item (Eval./Attrib. only):} \texttt{\{\{target\_text\}\}}
    
    \item \textbf{Generated reasoning trace:} \texttt{\{\{reasoning\_text\}\}}
\end{itemize}
\textbf{Output.} Return one JSON object. For each dimension in Table~\ref{tab:judge_dimensions}, provide \texttt{score} (an integer from 1 to 5) and a brief \texttt{reason}; include no other fields or markdown.
\end{minipage}%
}

\paragraph{Hard Scoring Constraints.}
To enforce factual faithfulness, the judge applies explicit score caps. If a trace claims clicks, conversions, searches, consultations, or lead submissions absent from the user history, its factuality score cannot exceed 2. Similarly, backfilling candidate or target information into the user history also limits factuality to at most 2. Fabricated locations, behavior times, consumption stages, or explicit user needs limit the score to at most 3, with lower scores assigned to severe violations. These constraints prioritize evidence-grounded reasoning over target consistency or explanatory completeness.

%% file: appendix_eval_tables.tex
\begin{table*}[t]
\centering
\caption{Task-specific requirements and general principles for LLM-based CoT quality evaluation.}
\label{tab:judge_task_principles}
\setlength{\tabcolsep}{2pt}
\renewcommand{\arraystretch}{1.38}
\setlength{\extrarowheight}{1.5pt}
\begin{tabular*}{\textwidth}{@{\extracolsep{\fill}}
    >{\RaggedRight\arraybackslash}p{0.15\textwidth}
    >{\RaggedRight\arraybackslash}p{0.32\textwidth}
    >{\RaggedRight\arraybackslash}p{0.45\textwidth}@{}}
\toprule
\textbf{Component} & \textbf{Definition} & \textbf{Evaluation Focus} \\
\midrule
\multicolumn{3}{@{}l}{\textit{Task-specific requirements}} \\
\addlinespace[2pt]

Exploration
& Infer multiple potential commercial intents from user profiles and historical behaviors without observing a target advertisement.
& Assess whether the trace identifies plausible and diverse intent directions, translates interests into serviceable demands, and provides evidence for the listed intents. \\

Evaluation
& Assess the match between the user evidence and an observed positive item.
& Assess whether the rationale justifies its 1--5 matching score from history, signal strength, user demand, and contextual constraints without fabricating evidence. \\

Attribution
& Explain an observed positive interaction.
& Assess whether the trace identifies plausible preceding evidence while avoiding post-hoc fabrication or unsupported certainty. \\

\midrule
\multicolumn{3}{@{}l}{\textit{General evaluation principles}} \\
\addlinespace[2pt]

Evidence priority
& Treat \texttt{history\_text} as the primary factual source and \texttt{user\_plain\_text} as supplementary information.
& Profile attributes and target information must not override, alter, or be presented as historical behavioral facts. \\

Behavior hierarchy
& Rank signals by conversion depth: conversion $>$ click $>$ exposure or viewing for more than three seconds.
& Repeated shallow exposures should not be treated as stronger intent than fewer clicks or conversions. \\

\makecell[l]{Fulfillment\\awareness}
& Distinguish between online and offline fulfillment.
& Offline services should consider location, service radius, accessibility, and demand timing; online services need not include geographic analysis unless a clear conflict exists. \\

\makecell[l]{Uncertainty\\awareness}
& Permit low matching scores and cautious attribution when the historical evidence is weak.
& Reward factual faithfulness and calibrated reasoning rather than forcing every signal into a strong explanation. \\

Anti-hallucination
& Prohibit fabricated clicks, conversions, searches, consultations, lead submissions, locations, behavior times, or explicit needs.
& Prioritize factuality over target consistency or explanatory completeness; traces must not invent evidence to support a target. \\
\bottomrule
\end{tabular*}
\end{table*}

\begin{table*}[t]
\centering
\caption{Scoring dimensions used by the LLM judge. Each dimension is rated on a five-point scale; representative low-, medium-, and high-quality anchors are provided.}
\label{tab:judge_dimensions}
\setlength{\tabcolsep}{2pt}
\renewcommand{\arraystretch}{1.24}
\setlength{\extrarowheight}{1.2pt}
\begin{tabular*}{\textwidth}{@{\extracolsep{\fill}}
    >{\RaggedRight\arraybackslash}p{0.14\textwidth}
    >{\RaggedRight\arraybackslash}p{0.26\textwidth}
    >{\RaggedRight\arraybackslash}p{0.18\textwidth}
    >{\RaggedRight\arraybackslash}p{0.18\textwidth}
    >{\RaggedRight\arraybackslash}p{0.18\textwidth}@{}}
\toprule
\textbf{Dimension}
& \textbf{Evaluation Criterion}
& \multicolumn{3}{c}{\textbf{Scoring Anchors}} \\
\cmidrule(l){3-5}
& & \textbf{Low (1)} & \textbf{Medium (3)} & \textbf{High (5)} \\
\midrule

Factuality
& Faithfulness to historical behaviors without fabrication, exaggeration, chronological confusion, or target backfilling.
& Substantial fabrication or severe detachment from history.
& Mostly based on history but contains local misreadings or overgeneralization.
& Fully grounded in history with no evident fabrication or backfilling. \\

Evidence selection
& Selection of the most relevant, strong, recent, and decision-relevant historical evidence.
& Relies on irrelevant evidence and ignores key signals.
& Uses partially relevant evidence but includes noise or misses competing signals.
& Accurately prioritizes key evidence and handles noise and competing directions. \\

Signal strength
& Correct weighting of conversions, clicks, and shallow exposures according to conversion depth.
& Treats weak exposure as decisive or ignores strong conversion evidence.
& Generally distinguishes signal levels but occasionally overweights frequency.
& Robustly weights behavior depth, direction concentration, and temporal order. \\

Intent translation
& Translation from content interests into local-service needs and actionable advertising intent.
& Remains entirely at the content-interest level.
& Identifies a service category but provides limited conversion context.
& Clearly identifies service needs, consumption scenarios, and potential conversion actions. \\

Causal logic
& Coherence of the chain from historical behavior to need, commercial intent, and final judgment.
& Contains severe logical gaps, contradiction, or forced causality.
& The main logic holds but some inference steps or stage transitions are weak.
& Forms a clear and well-supported reasoning chain with appropriate uncertainty. \\

Advertising fit
& Alignment with lead-ad recommendation rather than ordinary content recommendation.
& Only performs generic content matching.
& Identifies an advertising category but lacks conversion or decision factors.
& Characterizes service needs, conversion paths, and major commercial decision factors. \\

Spatiotemporal reasoning
& Appropriate handling of time, location, fulfillment mode, service radius, and accessibility.
& Ignores clear temporal or regional conflicts.
& Partially considers time or location but insufficiently distinguishes fulfillment modes.
& Applies restrained and appropriate spatiotemporal reasoning according to fulfillment type. \\

\makecell[l]{Confidence\\calibration}
& Consistency between conclusion strength or confidence and the quality of supporting evidence.
& Is clearly overconfident, underconfident, or internally inconsistent.
& Confidence is generally reasonable but locally insufficiently calibrated.
& Confidence accurately reflects evidence strength, logical completeness, and task uncertainty. \\

Task fulfillment
& Completion of the requirements associated with Exploration, Evaluation, or Attribution.
& Fails to perform the requested reasoning task.
& Completes the main task but omits a required intent list, matching score, attribution, or clear judgment.
& Fully follows the task-specific objective and required output structure. \\
\bottomrule
\end{tabular*}
\end{table*}